\documentclass[twocolumn]{aastex63}

\usepackage{enumitem}
\usepackage{mathtools}
\usepackage{csvsimple}
\usepackage{tabularx}
\usepackage[linesnumbered,ruled,vlined]{algorithm2e}

\shorttitle{PetroFit}
\shortauthors{Geda et al.}

\graphicspath{{./}{figures/}}

\begin{document}

\title{PetroFit: A Python Package for Computing\\ Petrosian Radii and Fitting Galaxy Light Profiles}

\author[0000-0003-1509-9966]{Robel Geda}
\affiliation{Space Telescope Science Institute STScI,  3700 San Martin Dr, Baltimore, MD 21218}

\author[0000-0002-8969-5229]{Steven M. Crawford}
\noaffiliation

\author[0000-0001-8587-9285]{Lucas Hunt}
\affiliation{United States Naval Observatory, 3450 Massachusetts Ave NW, Washington, DC 20392, USA}
\affiliation{Computational Physics, Inc., 8001 Braddock Road, Suite 210, Springfield, VA 22151-2110}

\author[0000-0002-3131-4374]{Matthew Bershady}
\affiliation{University of Wisconsin—Madison, Department of Astronomy, 475 N. Charter Street, Madison, WI 53706-1582, USA}
\affiliation{South African Astronomical Observatory, P.O. Box 9, Observatory 7935, Cape Town, South Africa}
\affiliation{Department of Astronomy, University of Cape Town, Private Bag X3, Rondebosch 7701, South Africa}

\author[0000-0002-9599-310X]{Erik Tollerud}
\affiliation{Space Telescope Science Institute STScI,  3700 San Martin Dr, Baltimore, MD 21218}

\author[0000-0001-5373-6669]{Solohery Randriamampandry}
\affiliation{South African Astronomical Observatory, P.O. Box 9, Observatory 7935, Cape Town, South Africa}
\affiliation{Southern African Large Telescope, P.O. Box 9, Observatory 7935, Cape Town, South Africa}
\affiliation{A\&A, Department of Physics, Faculty of Sciences, University of Antananarivo, B.P. 906, Antananarivo 101, Madagascar}

\published{AJ April 8, 2022}

\begin{abstract}
PetroFit is an open-source Python package, based on Astropy and Photutils, that can calculate Petrosian profiles and fit galaxy images. It offers end-to-end tools for making accurate photometric measurements, estimating morphological properties, and fitting 2D models to galaxy images. Petrosian metric radii can be used for model parameter estimation and aperture photometry to provide accurate total fluxes. Correction tools are provided for improving Petrosian radii estimates affected by galaxy morphology. PetroFit also provides tools for sampling Astropy-based models (including custom profiles and multi-component models) onto image grids and enables PSF convolution to account for the effects of seeing. These capabilities provide a robust means of modeling and fitting galaxy light profiles. We have made the PetroFit package publicly available on GitHub (\texttt{\href{https://github.com/PetroFit/petrofit}{PetroFit/petrofit}}) and PyPi (\texttt{pip install petrofit}). 
\end{abstract}


%% cross-references to Sections, equations, tables, and figures.

\section{Introduction} \label{sec:intro}

Since the advent of quantitative photometric studies of galaxies, measuring their true sizes and fluxes has proven to be a challenging task. This is because the true extent of a galaxy can be hard to define: it can vary with morphology, be difficult to distinguish from its neighbors, or difficult to measure due to low surface brightness features. Over time, researchers have proposed various approaches for estimating the fluxes, shapes, and angular (or projected) sizes of galaxies. A powerful technique for measuring accurate properties of galaxies is to combine parametric measurements with analysis of the profile of the light distribution of a galaxy.  

Parametric descriptions of galaxies were first introduced by  \cite{vaucouleurs1948}, who proposed a power-law profile that reasonably models the projected intensity profiles of galaxies. He used an $exp \{-k r^{1/4}\}$ intensity profile to describe the radial light distribution of elliptical galaxies. Following this model's success, \cite{sersic1963, sersic1968} introduced a generalized version that can be applied to galaxies of various morphologies.  Today, there are several parametric fitting codes, including GALFIT \citep{peng2010}, and statmorph \citep{rodriguezGomez2019MNRAS}.

Though the sizes of galaxies can be measured by fitting a Sérsic model to an image, various radii of interest can also be derived from radial photometric measurements.  \cite{petrosian1976ApJ} formulated a radial profile, now named after him, as a means of defining the projected radii of galaxies. Since the Petrosian profile is a ratio of surface brightnesses, it offers a distance-independent method of measuring galaxy radii. Petrosian radii are useful for computing radial concentrations of galaxy light profiles and for performing accurate measurements of  galaxy photometry \citep{graham2005PASA}. Petrosian radii and concentrations can also be used to estimate the parameters of Sérsic profiles. 

By combining parametric fitting and radial profile analysis, accurate fluxes, shapes, and sizes can be derived for studying the properties and evolution of galaxies.  In this paper, we introduce PetroFit,  a Python package based on the Astropy \citep{astropy13} and Photutils \citep{bradley2020} packages. The prime motivation for constructing this package was to make a robust and scalable software package for modeling Sérsic and Petrosian profiles. PetroFit is deigned to work within the Astropy ecosystem with a focus on measuring accurate galaxy properties.  Following \citep{crawford2006PhDT}, PetroFit uses correction grids to improve the size estimates produced by Petrosian profiles. PetroFit also provides a robust Python-based parameter fitter code based on Astropy models for parameterization of galaxy properties.  

%The Astropy project is an open-source effort to provide a free and accessible astronomical Python library. Photutils is a package based on Astropy that is focused on photometric measurements of various types of astronomical sources. 
%Upon recognizing PetroFit's potential, we have made the package publicly available with the hope of serving the astronomical community.

This paper  defines photometric terms and models used by PetroFit in Section \ref{sec:photo}. We outline the core features of the package in Section \ref{sec:software} and discuss how measurements are computed. We test and demonstrate various applications of the software in Section \ref{sec:testofsoftware}. We make qualitative comparisons to similar software available to the Astronomical community in Section \ref{sec:otherpackages}. Throughout this paper, the $AB$ magnitude system is used. Python Notebooks used to perform tests and generate plots can be found on GitHub in the \href{https://github.com/PetroFit/petrofit_papers}{\texttt{PetroFit/petrofit\_papers}} repository \citep{petrofitpapers}.

\section{Photometric Relations} \label{sec:photo}

In this Section, we define the morphological and photometric relations that can be computed using PetroFit. These properties can be best understood and categorized as parameters of Sérsic and Petrosian profiles, which we discuss in detail in the following subsections. We also introduce different techniques for accurate photometric measurements.

\subsection{Sérsic Profiles} \label{sersicprofiles}

The Sérsic profile \citep{sersic1963, sersic1968} is a  radial profile function that describes how the surface brightness of a galaxy varies with projected radius ($r$) from its center (see Figure \ref{fig:sersic}). The profile is a generalized form of the  de Vaucouleurs' profile \citep{vaucouleurs1948} and is often written as the following intensity profile:

\begin{equation} \label{eq:sersicprofile}
    I\left( r(x,y) \right) = I_e\exp\left\{-b_n\left[\left(\frac{r}{r_{e}}\right)^{(1/n)}-1\right]\right\}
\end{equation}

Where $I$ is the intensity at position $(x, y)$. $r$ is the radius from the center that corresponds to $(x, y)$. $r_{e}$ is the effective radius, which is equal to the projected half-light radius. $I_e$ is the intensity at the half-light radius ($I_e = I(r_{e})$). $n$ is the Sérsic index which determines the “steepness” of the profile. The constant $b_n$ is defined such that $r_{e}$ contains half the total flux. Figure \ref{fig:cog} shows an example Sérsic profile with $(n=1, I_e=1, r_e=25)$.

\subsubsection{Total Sérsic Flux}

The total flux (projected luminosity $L$) contained within a Sérsic profile over a projected area $A(r) = \pi r^2$ can be computed by integrating Equation \ref{eq:sersicprofile} as follows \citep{Ciotti1991}:

\begin{equation} \label{eq:fluxenclosedintegral}
    L(\leq r) = 2 \pi \int_{0}^{r} I(r^\prime) r^\prime dr^\prime
\end{equation}

\cite{graham1996ApJ, graham2005PASA} solve for $L(\leq r)$ by substituting $x = b_n (r / r_e)^{1/n}$ which yields:

\begin{equation} \label{eq:fluxenclosed}
    L(\leq r) = I_e r_{e}^2 2 \pi n \frac{e^{b_n}}{(b_n)^{2n}} \gamma (2n, x) 
\end{equation}

Where $\gamma (2n, x)$ is the incomplete gamma function:

\begin{equation} \label{eq:gamma}
    \gamma (2n, x) =  \int_{0}^{x} e^{-t} t^{2n-1} dt 
\end{equation}

Because the Sérsic profile is an analytical function, the total flux of the profile can be computed by integrating Equation \ref{eq:fluxenclosedintegral} to infinity. Substituting the complete gamma function $\Gamma(2n)$ in place of $\gamma (2n, x)$  in Equation \ref{eq:fluxenclosed} yields the total Sérsic flux \citep{Ciotti1991, graham2005PASA}:

\begin{equation}  \label{eq:totalsersicfluxenclosed}
    L(\leq \infty) = I_e r_{e}^2 2 \pi n \frac{e^{b_n}}{(b_n)^{2n}}  \Gamma(2n) 
\end{equation}

\subsubsection{Sérsic Index}\label{sersicindex}

The Sérsic index ($n$) determines the degree of curvature of the Sérsic profile. As the value of $n$ increases, the more concentrated the profile becomes at smaller radii. The Sérsic model tends to a power-law with a slope equal to 5 as $n$ becomes larger \citep{graham2005PASA}. $n = 4$ corresponds to a de Vaucouleurs' profile that describes elliptical galaxies and the cores of spirals well; while $n = 1$ gives the exponential profile, which models spiral disks well \citep{trujillo2001, andredakis1995MNRAS, khosroshahi2000ApJ, graham2001AJ, mollenhoff2001AA, moriondo1998AA}. $n = 0.5$ gives a low concentration profile (Gaussian) that can be used to describe seeing dominated images, model intra-cluster glow in galaxy clusters (assuming an elliptical shape), and is often used as an alternative to a Ferrer Profile \citep{peng2010}. 

% \begin{figure}
% \includegraphics[width=8cm]{sersic_index.png}
% \centering
% \caption{A two dimensional  plot of Sérsic indices that produce Gaussian ($n = 0.5$), exponential ($n = 1$) and de Vaucouleurs' ($n = 4$) profiles. The Sérsic profiles are centered at $r=0$ and the plot ranges from $-100$ to $100$. Notice the increased concentration towards the center of the profile with increasing Sérsic index.}
% \end{figure}

\subsubsection{Effective Radius and Effective Intensity}

The effective ($r_{e}$) or half-light ($r_{50}$) radius is the radius that encloses half the total flux of the Sérsic profile \citep{Ciotti1991}. As such, we define $r_{e}$ as the radius that satisfies the following expression:

\begin{equation}\label{eq:redef}
    L(\leq r_e) = \frac{1}{2}L(\leq \infty)
\end{equation}

Note that the expression given by Equation \ref{eq:redef} reduces to Equation \ref{eq:gammarelation}. Thus $b_n$ is defined as the value that satisfies the following relation between the complete and incomplete gamma functions \citep{Ciotti1991}:

\begin{equation}\label{eq:gammarelation}
    \gamma (2n, b_n) = \frac{1}{2}\Gamma(2n)  
\end{equation}

The effective intensity ($I_e$) is the intensity exactly at $r_{e}$ and can be defined as $I_e = I(r_{e})$. $I_e$ determines the amplitude of the profile and it is related to the intensity at the center of the Sérsic profile ($I_0 = I(0)$)  as follows:

\begin{equation}
    I_e = \frac{I_0}{\exp\left\{b_n\right\}}
\end{equation}

\begin{figure} 
\includegraphics[width=8cm]{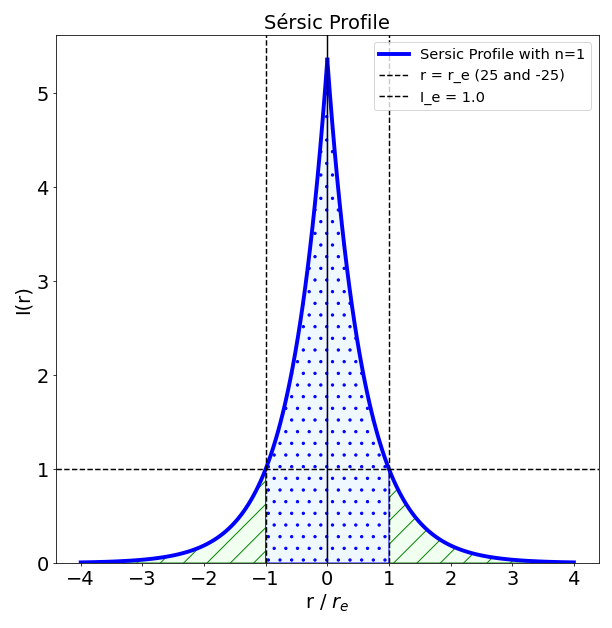}
\centering
\caption{\label{fig:sersic}
A unitless one dimensional Sérsic profile is plotted with $n = 1$, $r_e = 25$ (green vertical lines), and $I_e = 1$ (black horizontal line). The flux at the center of the profile ($I_0$) is equal to 5.3 and not 1 (i.e. $I_0 \neq I_e $). The dotted blue area under the curve contains half of the total flux and is equal in value to the striped green area (that also contains half of the total flux). The Sérsic profile and the green area extend to infinity.
}

\vspace{5mm}

\includegraphics[width=8cm]{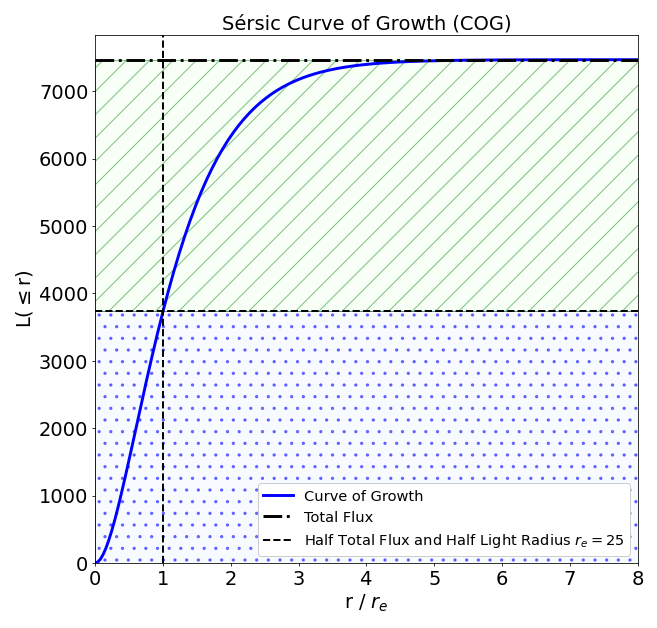}
\centering
\caption{\label{fig:cog}
The curve of growth (COG) represents the total flux enclosed at each radius and asymptotically approaches the total flux (black dash-dot horizontal line). This example shows the curve of growth of a Sérsic profile with $(n=1, I_e=1, r_e=25)$. The half-light radius ($r_e = 25$) contains half of the total flux that is contributed by the blue dotted area in Figure \ref{fig:sersic}.
}
\end{figure}

\subsubsection{Ellipticity, Elongation, and Orientation}

Real galaxies, that are well described by Sérsic profiles, are usually not perfectly symmetrical and often have elliptical distributions \citep{ferrari2004}. We define two quantities, ellipticity ($ellip$) and elongation ($elong$), that can be used to describe the elliptical distribution of Sérsic profiles:

\begin{equation}
    ellip =  1  -  \frac{b}{a} = \frac{elong - 1}{elong}
\end{equation}

\begin{equation}
    elong = \frac{a}{b} = \frac{1}{1 - ellip}
\end{equation}

Where $a$ is the unit length of the semi-major axis and $b$ is the corresponding semi-minor axis. A circle corresponds to $ellip = 0$ and $elong = 1$. Ellipticity ranges from 0 to 1 while elongation ranges from 1 to infinity. The SExtractor \citep{sextractor1996} and  
\href{https://photutils.readthedocs.io/en/stable/api/photutils.segmentation.SourceProperties.html\#photutils.segmentation.SourceProperties.elongation}{Photutils} packages use elongations for apertures while the Astropy-modeling sub-module uses ellipticity for \href{https://docs.astropy.org/en/v4.2/api/astropy.modeling.functional_models.Sersic2D.html}{Sérsic models}. 

Ellipticities ($ellip$) and rotation angles ($\theta$) are used to relate Cartesian coordinates to elliptical radii by the Astropy and PetroFit packages. For an elliptical profile centered at $(x_0, y_0)$, with an effective radius $r_e$, the following relation is used \footnote{\href{https://docs.astropy.org/en/v4.2/_modules/astropy/modeling/functional_models.html\#Sersic2D.evaluate}{See Astropy's (v4.2) \texttt{Sersic2D.evaluate} function}}:

\begin{equation}
    r(x, y) = \sqrt{\left(r_{maj}(x, y)\right)^2 - \left(\frac{r_{min}(x, y)}{(1 - ellip)}\right)^2}
\end{equation}

Where:

$r_{maj}(x, y) = (x - x_0)\cdot{cos(\theta)} + {(y - y_0)}\cdot sin(\theta)$

$r_{min}(x, y) = -(x - x_0)\cdot{sin(\theta)} + {(y - y_0)}\cdot cos(\theta)$

\subsubsection{Curve of Growth}

The curve of growth (COG) of a galaxy is the relationship between radius (from the galaxy center) and the total intensity within that radius. It represents the cumulative flux enclosed at a given radius. For an ideal galaxy that is well described by a Sérsic profile, the curve of growth is given by $L(\leq r)$. Figure \ref{fig:cog} shows an example curve of growth for a Sérsic profile with $(n=1, I_e=1, r_e=25)$.

\subsection{Petrosian Profiles}\label{sec:petrosian}

The Petrosian profile or Petrosian index ($\eta$) is a dimensionless profile that represents the rate of change in the enclosed light as a function of radius. It was first introduced by \cite{petrosian1976ApJ} who conceptualized it as a means of defining the sizes (radii) of galaxies. Petrosian originally defined the Petrosian index as a ratio of the average surface brightness up to a radius divided by the surface brightness at that radius. Since $\eta$ is defined as an intensity ratio, it is not affected by the surface brightness dimming effects that result from the cosmological expansion of the Universe. PetroFit follows \cite{kron80, bershady2000AJ} and defines the Petrosian profile ($\eta$) as the reciprocal of the original formulation by Petrosian. This results in a profile that has the property where $\eta(0) = 1$ at the center of the galaxy and drops to zero at the edge. As such, the Petrosian profile ($\eta(r)$) is defined as:

\begin{equation} \label{eq:petrosian}
    \eta (r) = \frac{I(r)}{\langle I(r) \rangle} = I(r) \frac{A(r)}{L(\leq r)}
\end{equation}

Where $\eta (r)$ is the Petrosian index at $r$. $I(r)$ is the surface brightness at $r$. $\langle I (r) \rangle$ is the average surface brightness within $r$. $L(\leq r)$ is the total flux within $r$. $A(r)$ is the aperture area. For elliptical apertures, the area is given as $A(r) = \pi \cdot (1 - ellip) \cdot r^2$.

\cite{graham2005PASA, grahamPetrosian2005AJ} derive the Petrosian profile of a Sérsic profile as a function of the logarithmic slope of the Sérsic profile $\alpha (x, n)$ \footnote{\cite{grahamPetrosian2005AJ} uses the original definition of the Petrosian profile, thus the expression they provide is the reciprocal of Equation \ref{eq:etaofxn}} :

\begin{equation} \label{eq:etaofxn}
    \eta (x, n) = \frac{2}{\alpha (x, n)} = \frac{2 n \gamma(2n, x)}{e^{-x} x^{2n}}
\end{equation}

Where $x$ is given by the function  $x(r) = b_n (r / r_e)^{1/n}$, $n$ is the Sérsic index and $ \gamma$ is the incomplete gamma function.

\begin{figure}
\includegraphics[width=8cm]{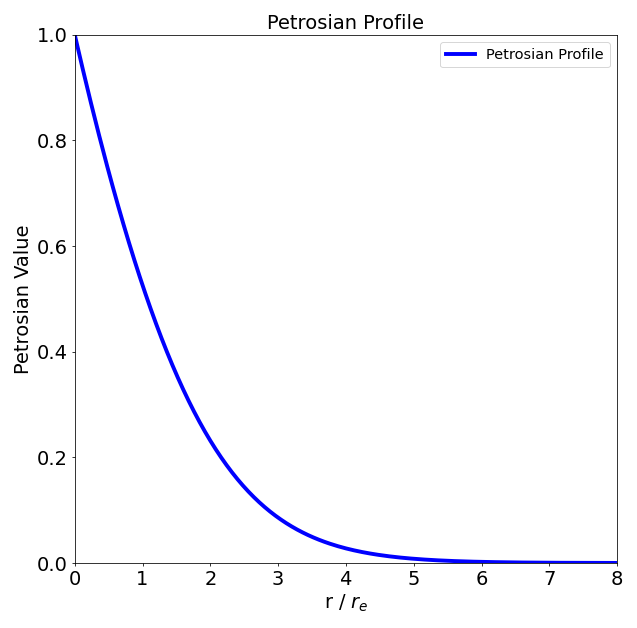}
\centering
\caption{A plot of the Petrosian profile ($\eta (r)$) of a Sérsic profile with $(I_e=1, r_e=25,  n=1)$. Note that that the profile has the property where $\eta(0) = 1$ at the center of the galaxy ($r=0$) and drops to zero at the edge.}
\end{figure}

\subsubsection{Petrosian Radius and Eta} \label{petrosianeta}

The Petrosian radius is the radius where the Petrosian profile is equal to a special Petrosian index  ``eta" ($\eta_{petro}$). Following \cite{bershady2000AJ}, the Petrosian radius is defined as the radius where $\eta = 0.2$. Thus the Petrosian radius ($r_{petro}$) and ``eta" ($\eta_{petro}$) can be expressed as:

\begin{equation}
    \eta_{petro} \equiv \eta_{0.2} = \eta(r_{petro}) = 0.2
\end{equation}

\subsubsection{Petrosian Total Flux Radius and Epsilon} \label{petrosianepsilon}

The Petrosian total flux radius is, as the name implies, the radius that encloses the total flux of the galaxy. Since Sérsic profiles extend to infinity, it is not possible to define a radius that contains $100\%$ of the total flux (that would equate to a radius at infinity). It is also not practical to consider minute flux values at very large radii because real images of galaxies contain noise. For these reasons, PetroFit considers the radius that contains $99\%$ of the galaxy’s light to be the Petrosian total flux radius \citep{crawford2006PhDT}. To calculate the Petrosian total flux radius, we must multiply the Petrosian radius by a constant. This constant is called ``epsilon" and is set to 2 by default \citep{bershady2000AJ}. This default value is also adopted by the Sloan Digital Sky Survey \citep{strauss2002AJ}. Thus we can define the Petrosian total flux radius as follows:

\begin{equation}
    r_{total} = \epsilon \cdot r_{petro}
\end{equation}

\subsection{Total Flux and Petrosian Corrections}\label{totalfluxdef}

 The Petrosian Total Flux is computed by setting $\epsilon=2$ by default (see Section \ref{petrosianepsilon}). But this approach is an approximation and is susceptible to errors that arise from different galaxy morphologies. In particular, $\epsilon$ is dependent on how concentrated a galaxy's light profile is. The more concentrated a galaxy's light profile is, the larger the value of $\epsilon$ needed to produce $r_{99}$. To understand this, let us consider two profiles, A and B, with the same total flux radii (i.e equal $r_{99}$), but with different concentrations (i.e $n(A) > n(B)$). Since profile A has a higher concentration, it will have a $r_{petro}$ closer to the center of the profile relative to $r_{99}$ simply because its profile is more concentrated. Thus the multiplication factor ($\epsilon$) needed to produce the total flux radius ($r_{99}$ or $r_{total}$) for A is relatively larger than B. Thus, to accurately estimate $r_{99}$, we must apply corrections by first estimating the appropriate $\epsilon$ value. Once the appropriate $\epsilon$ value is estimated, $r_{total}$ (which PetroFit defines as $r_{total} \equiv r_{99}$) can be computed by $r_{total} = r_{petro} \cdot \epsilon$. 
 
\subsubsection{Petrosian Half-Light Radius}

The Petrosian half-light radius is the radius that contains half of the Petrosian total flux. This quaintly is especially important because it is approximately (exactly in ideal cases) equal to the Sérsic effective radius. The Petrosian half-light radius can be computed by numerically finding the radius that encloses half of the Petrosian total flux. Consequently the Petrosian half-light radius ($r_{50}$) can be defined as the radius that satisfies the following expression:

\begin{equation}
    L(\leq r_{50}) = \frac{1}{2} L(\leq r_{total})
\end{equation}

\subsubsection{Concentration Index} \label{concentrationindex}

The concentration index is the ratio of the radii containing a specified fraction of the total flux and can be generalized as follows \citep{bershady2000AJ, kent1985ApJS}:  

\begin{equation}
    C_{i o} \equiv 5 \cdot log\left(\frac{r_{o}}{r_{i}}\right)
\end{equation}

where $r_o$ and $r_i$ are the outer and inner radii, enclosing $o\%$ and $i\%$ percent of the total flux respectively. The concentration index is related to the profile shape and can be used as a proxy for morphology \citep{crawford2006PhDT, grahamgrwincaon2001ApJ}. Some commonly used concentration indices are $C_{2080}$ \citep{bershady2000AJ, Conselice2003ApJS, kent1985ApJS} and $C_{5090}$ \citep{blanton2001AJ}:

\begin{equation}
    C_{2080} = 5 \cdot log\left(\frac{r_{80}}{r_{20}}\right)
\end{equation}

\begin{equation}
    C_{5090} = 5 \cdot log\left(\frac{r_{90}}{r_{50}}\right)
\end{equation}

$C_{2080}$ correlates better with the Sérsic index (and epsilon). However, in the presence of image degradation, $C_{5090}$ may be more immune from systematic effects.

\subsubsection{Petrosian Concentration Index} \label{petrosianconcentrationindex}

We define the concentration of radii derived from Petrosian indices as follows: 

\begin{equation}
    P_{i o} \equiv 5 \cdot log\left(\frac{r(\eta_{o})}{r(\eta_{i})}\right)
\end{equation}

where $r(\eta_{0})$ and $r(\eta_{i})$ are the outer and inner Petrosian radii, corresponding to $\eta_{0}$ and $\eta_{i}$ Petrosian indices respectively. This definition of concentration is useful because it can be computed directly from the Petrosian profile, without the need to estimate radii using the curve of growth (see Appendix \ref{epsilonvsp}). 

We also define a Petrosian concentration index, $P_{0502}$,  as follows:  

\begin{equation}
    P_{0502} = 5 \cdot log\left(\frac{r(\eta_{0.2})}{r(\eta_{0.5})}\right) = 5 \cdot log\left(\frac{r_{petro}}{r(\eta_{0.5})}\right) 
\end{equation}

\section{Software} 
\label{sec:software}

PetroFit is a Python-based package for computing and fitting various galaxy profiles. It builds on packages such as Astropy and Photutils to provide a set of tools for end-to-end profile construction and parameter fitting. As an open-source package, PertroFit is made available to the public and can be installed via PyPi. We discuss some of PetroFit's key features in this Section.

\subsection{Photometry Tools}

In order to make accurate measurements of galaxy properties, we need to make accurate photometric measurements while processing images of galaxies. Photutils \citep{bradley2020}, an Astropy affiliated package for photometry, offers a wide array of tools to perform photometry. We modify or build on the tools in Photutils, and in this subsection, we introduce some of the photometry utilities we have made available in PetroFit.  

\subsubsection{Image Segmentation and Source Catalogs}

It is often necessary to identify and make a catalog of sources in an image before making photometric measurements. It is also useful to identify  pixels in the image that are associated with a given source so a mask can be constructed for that source. This is used to exclude the light from neighboring objects that would otherwise contaminate or bias the light profile. The PetroFit \texttt{make\_catalog} function uses photometry routines in Photutils to make segmentation maps, deblend sources that overlap in the image, and produce a catalog with useful photometric information about the sources.

To identify sources, the image is partitioned into multiple segments and a segmentation map is constructed. To achieve this, the image is first smoothed with a Gaussian kernel if a kernel size is provided and segmented using Photutils' \texttt{detect\_sources} functionality. Photutils constructs a segmentation map by assigning a label to every pixel in the smoothed image such that pixels with the same label are part of the same source. A minimum source size can be specified by providing a cutoff for the number of connected pixels that have values above a user-provided threshold. Overlapping sources are separated using Photutils' \texttt{deblend\_sources} function that uses a combination of multi-thresholding and watershed segmentation. The resulting catalog and segmentation map can then be used to identify sources of interest and mask pixels of nearby sources when performing photometry measurements. PetroFit also provides plotting tools to visualize Photutils segmentation maps.

\subsubsection{Background Noise Subtraction} \label{backgroundNoiseSubtraction}

To construct a curve of growth that converges to the total flux of a galaxy, we must carefully estimate the noise in the image and make the appropriate subtractions. This is especially important for faint galaxies that have low signal-to-noise ratios. To deal with this, PetroFit offers an  ``image continuum" fitting tool, \texttt{fit\_background}, that fits background pixels using an Astropy model, and then that model can be subtracted from the image. 

The background subtraction tool first makes a background image by masking all the sources identified in the segmentation stage. Note that our definition of background here is pixels that do not belong to a source. Once the background image is made, background pixels above a specified threshold value are also masked. The remaining unmasked pixels are used to fit an Astropy model. The default model is set to a  \href{https://docs.astropy.org/en/v4.2/api/astropy.modeling.functional_models.Planar2D.html}{Astropy \texttt{Planar2D} model}. 

The plane model is fit using a linear least-square  fitting algorithm provided by Astropy’s modeling module and directly sampled (without integrating or oversampling) into a two-dimensional model image. The user can then subtract the background image from the initial image to produce a background-subtracted image.

\subsubsection{Aperture Photometry}

The curve of growth and the Petrosian $\eta$ function are both radial profiles. As such,  photometric measurements must be made using apertures of varying radii from the center of the source. It is also important to  track the aperture areas and account for masked pixels since the Petrosian is computed using average surface brightness (see Equation \ref{eq:petrosian} and Section \ref{discretepetrosian}).

The PetroFit \texttt{photometry\_step} function constructs multiple concentric elliptical apertures and makes accurate flux measurements using Photutils. The elliptical shape of the apertures can be set by providing an elongation and rotation angle (default values correspond to a circle). The radii of the apertures are defined by the user and provided as a python list or array. PetroFit offers utility functions that can make lists of linearly or logarithmically spaced radii. Error images and mask maps can be provided for error estimation and pixel masking, respectively. The method returns the result in a tuple of three separate arrays that contain the flux measurements, the error estimates, and aperture areas. This function can also over-plot the apertures used for the measurement on the input image. 

The PetroFit \texttt{source\_photometry} function is designed to work with Photutils source catalogs and segmentation maps. Given a source in the Photutils catalog, the function masks surrounding pixels that belong to nearby sources (each source pixel is assigned to a single source),  optionally performs background subtraction using a two-dimensional plane model and then runs the \texttt{photometry\_step} function. The \texttt{source\_photometry} function also determines the appropriate aperture shape (i.e. elongation or ellipticity) and orientation using the information in the Photutils catalog.

\subsection{Petrosian Tools}

Given photometric measurements, PetroFit can compute the Petrosian profile as well as associated radii and concentrations. In this Section, we discuss how the Petrosian $\eta(r)$ is discretely computed and how galaxy properties can be derived from it. We also discuss corrections applied to the measurements to provide accurate photometry. 

\subsubsection{Analytically Computed Profiles}\label{analyticalprofiles}

PetroFit includes functions to compute Petrosian profiles and curves of growth analytically for ideal Sérsic profiles with known parameters. The following profiles and utilities are available as part of the package:

\begin{description}[leftmargin=0pt]

\item [\texttt{\bf \boldmath petrosian\_profile ($\eta (r)$)}] Given the effective radius and Sérsic index of a Sérsic profile, this function computes the Petrosian profile using Equation \ref{eq:etaofxn}. A 1D Astropy fittable model version is also available for this function but may not be useful for fitting profiles derived from real galaxy images (see Section \ref{petrosiancorrections}). 

\item [\texttt{\bf \boldmath sersic\_enclosed ($L(\leq r)$)}] Given effective intensity, effective radius and Sérsic index of a Sérsic profile, this function computes the total enclosed flux at an input radius according to Equation \ref{eq:fluxenclosed}. This function can be used to make a curve of growth for ideal Sérsic profiles. A 1D Astropy fittable model version is also available for this function.

\item [\texttt{\bf \boldmath sersic\_enclosed\_inv ($r(L)$)}] Computes the inverse of Equation \ref{eq:fluxenclosed} by solving for $r$. Given effective intensity, effective radius and Sérsic index of a Sérsic profile, this function computes the radius that encloses the input fraction of the total light flux. 
\end{description}

\subsubsection{Discretely Computed Petrosian Profiles} \label{discretepetrosian}

Given an array of enclosed fluxes (``$L$") and corresponding aperture areas (``$A$"), the Petrosian profile can be computed discretely as follows:

1. Estimate the surface brightness by finding the average flux between the current index ($i$) and the last index ($i-1$). Note that the gap between apertures affects the accuracy of the surface brightness at $i$, thus it is recommended to use apertures with radii that are incremented by a small number of pixels:

\begin{equation}
    I[i] \approx \frac {L[i] - L[i-1]} {A[i] - A[i-1]}
\end{equation}

2. Estimate the average surface brightness by taking the flux at $i$ and dividing it by the corresponding aperture area:

\begin{equation}
    {\langle I[i] \rangle} = \frac {L[i]} {A[i]}
\end{equation}

3. Compute the Petrosian index at $i$ using the estimated values in steps 1 and 2:

\begin{equation}
    \eta [i] = \frac{I[i]}{\langle I[i] \rangle} = \left(\frac {L[i] - L[i-1]} {A[i] - A[i-1]}\right) \cdot \frac {A[i]} {L[i]}
\end{equation}

In discrete computations, the Petrosian profile can not be computed at the first index even if it corresponds to the center pixel ($r[i_0]=0$).  In real images, the surface brightness of a galaxy is binned into pixels, and to accurately determine $I_0$, one would need to infinitely oversample the central region. In other words, each pixel corresponds to a total surface brightness integrated within the area of the pixel as opposed to the surface brightness at the pixel coordinates. As such, PetroFit sets the first Petrosian value to \texttt{numpy.nan} when returning a discretely computed array of Petrosian indices. PetroFit takes advantage of fact that the Petrosian index at the center of a galaxy is equal to $1$ when computing radii internally.

\subsubsection{Petrosian Radii and Concentration Indices}

Given a curve of growth measurement in the form of arrays containing aperture radii, aperture areas, and enclosed fluxes, PetroFit constructs a Petrosian profile as described in Section \ref{discretepetrosian}. Once the Petrosian profile is constructed, Petrosian radii and concentrations can be computed. We describe how each property is computed in this Section.  

\begin{description}[leftmargin=0pt]

\item [\texttt{\bf \boldmath r\_petrosian ($r_{petro}$)}] The Petrosian radius is computed by identifying the radius where the Petrosian profile reaches the Petrosian eta ($\eta_{petro}$) value. By default, $\eta_{petro}$ is set to $0.2$ as discussed in Section \ref{petrosianeta}. The values of the Petrosian profile are interpolated and the radius where the profile intersects with $\eta_{petro}$ is identified as $r_{petro}$. The eta value can be user adjusted and interpolation can be switched off (to find the closest data-point) if necessary. 

\item [\texttt{\bf \boldmath r\_total\_flux ($r_{total}$)}] The total flux radius is computed by multiplying \texttt{r\_petrosian} with epsilon (see Section \ref{petrosianepsilon}). Epsilon is set to 2 by default and can be adjusted. $r_{total}$ depends on the assumed functional form of the radial profile of the galaxy and we discuss how to make these corrections in Section \ref{petrosiancorrections}. If the image includes a World Coordinate System, PetroFit will convert the radius into arcsec units based on the corresponding pixel scale.

\item [\texttt{\bf \boldmath total\_flux ($ L(\leq r_{total})$)}] The total Petrosian flux is computed by referring to input curve of growth and identifying the total enclosed flux at \texttt{r\_total\_flux}. If the curve of growth does not extend to the \texttt{r\_total\_flux}, a Numpy \texttt{nan} is returned. Note that one can set \texttt{r\_total\_flux} to the largest aperture radius available in the photometric data ($r_{max}$) by defining the appropriate epsilon value (i.e $\epsilon = r_{max} / r_{petro}$).    

\item [\texttt{\bf \boldmath total\_flux\_uncertainty}] The uncertainty or error in the total enclosed flux at \texttt{r\_total\_flux}. Only available if photometric uncertainties (\texttt{astropy.uncertainty} object) are provided.     

\item [\texttt{\bf \boldmath r\_half\_light ($r_{50} $)}] The half-light radius is computed by finding the radius that encloses half of \texttt{total\_flux} in the curve of growth (see Figure \ref{fig:sersic} and \ref{fig:cog}). This value can be converted into arcsec by providing a WCS.

\item [\texttt{\bf \boldmath concentration\_index ($C_{io}$)}] The concentration index is computed by first identifying the the radii that contain the requested fractions of the total flux and computing their concentration ratio as described in Section \ref{concentrationindex}.

\end{description}

\subsubsection{Petrosian Corrections}\label{petrosiancorrections}

\begin{figure} 
\centering

\includegraphics[width=8cm]{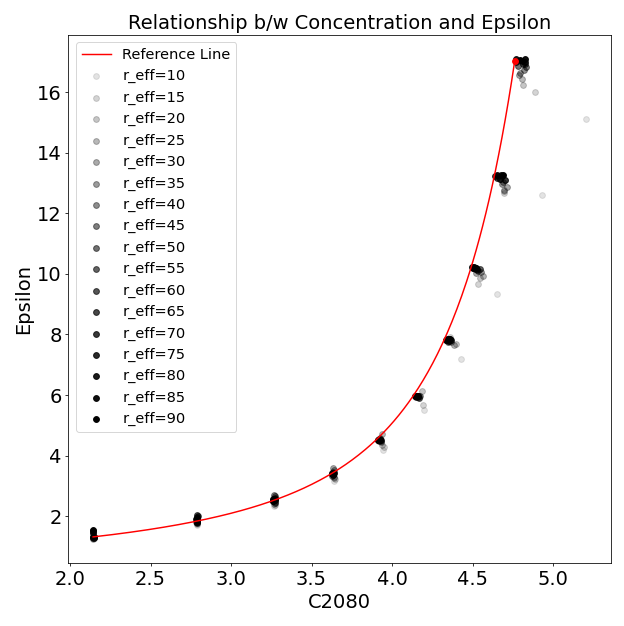} 

\caption{\label{fig:concentrationvsepsilon}
In this figure, we plot epsilon ($\epsilon$) vs. $C_{2080}$ (uncorrected) for simulated 2D Sérsic profiles with varying effective radii (black points). The scatter seen at higher concentrations is due to sampling errors. The red curve is a line of best fit given by Equation \ref{eq:c_vs_epsilon_approx} in Appendix \ref{epsilonvsc}. The $C_{2080}$ domain corresponds to $0.5 \leq n \leq 5$ in Sérsic index. 
}

\vspace{4mm}

\includegraphics[width=8cm]{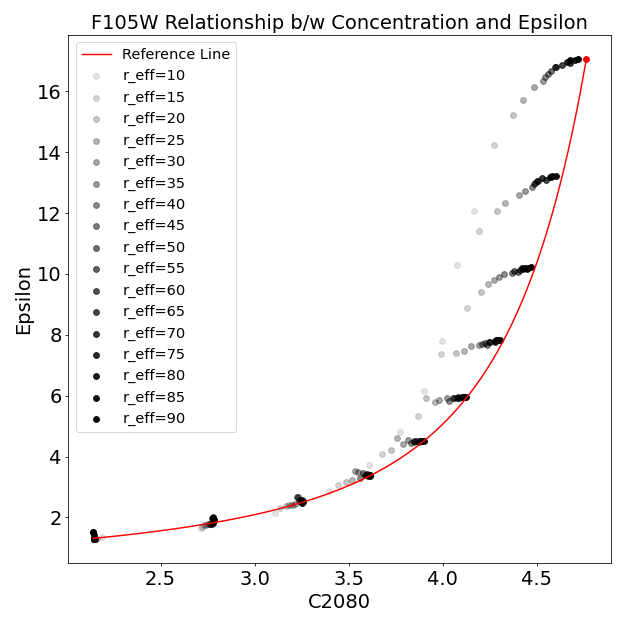}
\centering
\caption{\label{fig:concentrationvsepsilonPSF}
This figure illustrates PSF effects by plotting $\epsilon$ vs. $C_{2080}$ (uncorrected) for the same profiles as Figure \ref{fig:concentrationvsepsilon}, but with the profiles convolved with a HST WFC3 F105W PSF (sampled at $60 mas$). It also shows the same curve as Figure \ref{fig:concentrationvsepsilon} for reference. Notice how $\epsilon$ and $C_{2080}$ values of profiles with smaller effective radii are affected (skewed) the most due to the PSF convolution. For this reason, we must make corrections that account for effective radii and the shape of the PSF itself. 
}
\end{figure}

To first order, the Petrosian magnitude is independent of redshift in the local Universe as it depends on the ratio of two surface brightnesses, but because it is based on profile shape, changes in the profile due to morphological K-corrections still apply \citep{crawford2006PhDT}. However, the real strength of Petrosian magnitudes vs. other variable aperture photometry such as Kron magnitudes \citep{kron80} is that the Petrosian radius only depends on the light interior to this radius. This aspect leads to small random errors in the measurement of $r_{petro}$. Nonetheless, the magnitude within $2 \cdot r_{petro}$, relative to a true, total magnitude, is still profile-dependent. Although $99\%$ of the light from an exponential profile ($n=1$) is recovered within $2 \cdot r_{petro}$, only $82\%$ of the light for a $r^{1/4}$ profile is measured within $2 \cdot r_{petro}$ \citep{graham2005PASA}. Similar results apply to the Kron \citep{kron80} scheme, as discussed in \cite{Bershady1995AJ}. Because of this, we need to adjust $r_{total}$ by finding the appropriate epsilon ($\epsilon$) value (see Section \ref{petrosianepsilon}).

The concentration index and  ``correct" $r_{total}$ are related, thus a relationship between the concentration index and $\epsilon$ can be derived. We attempt to derive this relationship by simulating  Sérsic profiles with varying concentrations and measuring the radii that contain $99\%$ of the total fluxes (i.e. $r_{total}$, see Appendix \ref{paramapproximationss}). Real images of galaxies contain distortions through passing through the atmosphere or the optics of a telescope, and the images are smeared according to the point spread function (PSF). The smearing of light results in the reduction in profile concentration, causing a skew in the relationship derived from simulated profiles  (see Figures \ref{fig:concentrationvsepsilon} and \ref{fig:concentrationvsepsilonPSF}). To accurately calculate the total flux, further corrections must be applied to the estimates of $\eta$ based on the PSF, concentration, and Petrosian radius. To make this correction, PetroFit uses a pre-generated grid of corrected and uncorrected Petrosian radii, eta values, Sérsic indices, and concentrations. It uses the grid to apply appropriate corrections based on raw uncorrected measurements. To generate the grid, a Sérsic profile of known parameters is first generated and convolved by a PSF provided by the user. Once the image is convolved with the PSF, Petrosian measurements are made and stored in a table along with the known correct parameters. This allows us to also derive relationships between the Petrosian and Sérsic parameters such as concentration indices and Sérsic indices.

\subsection{Model Fitting Tools}

PetroFit's fitting module contains useful tools to model and fit a wide variety of galaxy types. It uses machinery provided by Astropy to model data such as background gradients as described in Section \ref{backgroundNoiseSubtraction}. Another useful feature of PetroFit is that it can fit the light profiles of galaxies by constructing Astropy supported models or a combination of them (compound models). We discuss how PetroFit fits galaxy light profiles in detail in this Section.

\subsubsection{Astropy Models}

The Astropy \texttt{modeling} module provides a framework for representing models and parameter fitting. Several predefined 1-D and 2-D models are available along with the ability to define custom models (See Appendix \ref{apdx:listofmodels}). Different fitting algorithms can be used depending on the model. Astropy’s \texttt{Sersic2D} model implements the Sérsic model as described in Section \ref{sersicprofiles} and allows users to provide initial guesses for the Sérsic parameters. Making good estimates of these parameters is important because the Levenberg-Marquardt algorithm is sensitive and may return parameters that correspond to a local (rather than global) minimum. Astropy’s  \href{https://docs.astropy.org/en/v4.2/api/astropy.modeling.functional_models.Sersic2D.html}{\texttt{Sersic2D} model} has the following parameters:

\begin{description}[leftmargin=0pt]
\item [\texttt{amplitude}] Surface brightness at r\_eff ($I_e$).
\item [\texttt{r\_eff}] Effective (half-light) radius ($r_e$).
\item [\texttt{n}] Sérsic Index, corresponds ($n$).
\item [\texttt{x\_0 \& y\_0}] center x and y position ($x_0, y_0$).
\item [\texttt{ellip}] Ellipticity of the profile, corresponds to ($ellip$).
\item [\texttt{theta}] Rotation angle in radians, counterclockwise from the positive x-axis ($\theta$).
\end{description}

\subsubsection{PetroFit PSFModel} \label{petroFitPSFModel}

PetroFit’s \texttt{PSFModel}\footnote{As of PetroFit version 0.4.0, PetroFit’s “PSFModel” class has been renamed to “PSFConvolvedModel2D” to distinguish it from and avoid confusion with Photutil’s PSF modeling library.} is a 2D image wrapper model that can be fit using Astropy's fitting algorithms. \texttt{PSFModel} is a subclass of Astropy’s \texttt{Fittable2DModel} and converts regular Astropy 2D models into images before each fitting iteration. It can wrap any Astropy based 2D model (including compound models) and inherits the base model's parameters. The base model can be retrieved using \texttt{PSFModel.model}, which would return the base model but with the current \texttt{PSFModel} parameters. During the fitting process or at evaluation, \texttt{PSFModel} samples the base model and translates it into an image via the following steps:

1. During initialization, \texttt{PSFModel} is given a base model, a normalized PSF image, and an oversampling parameter. \texttt{PSFModel} inherits the base model's parameters as well as its fixed-parameter rules and boundaries. 

2. A sampling grid is generated according to the oversampling parameter. The generated grid is stored and is only regenerated if the oversampling factor is changed or the oversampling parameter is a function of the model parameters. 

3. A model image is constructed using the sampling grid from the step before. To achieve this, the base model is evaluated onto the grid using the input parameters provided by the user or fitter. 

4. If the model was oversampled, the model image or oversampled region is mean block-reduced to the data resolution. See the Section \ref{oversampling} for more details.

5. The model image is convolved with the input PSF and the result is returned. The PSF should be normalized and have the same pixel resolution as the image. The \texttt{PSFModel} has a parameter \texttt{psf\_p}, which controls the rotation angle of the PSF image before convolution, in addition to the parameters from the base model. Using \texttt{psf\_p}, the fitter can rotate the PSF to produce a better result (lower residuals). Enabling \texttt{psf\_p} can be useful for concentrated sources or star-like objects that are being fit using PSFs obtained from a separate data source (diffraction spikes may not align). This feature can be disabled by adding \texttt{psf\_p} to the model's dictionary of fixed parameters. If a PSF is not provided, the model image is returned without further processing.

\subsubsection{Oversampling} \label{oversampling}

Visible and near-visible photons from galaxies are measured using optical telescopes and imaging devices such as CCDs or CMOS cameras. Unlike intensity models which enjoy infinite angular resolutions, images are composed of pixels that are limited in resolution. Since we are using images of galaxies as our fitting data sets, we sometimes need to make sure that analytical models expressed in intensity (surface brightness) are sampled onto pixels (fluxes) with the appropriate integration of the model. Because integrating such intensity models over each pixel's angular footprint is computationally costly (especially for fitting routines), PetroFit employs oversampling schemes to discretely integrate the models.

\begin{figure}
\includegraphics[width=8cm]{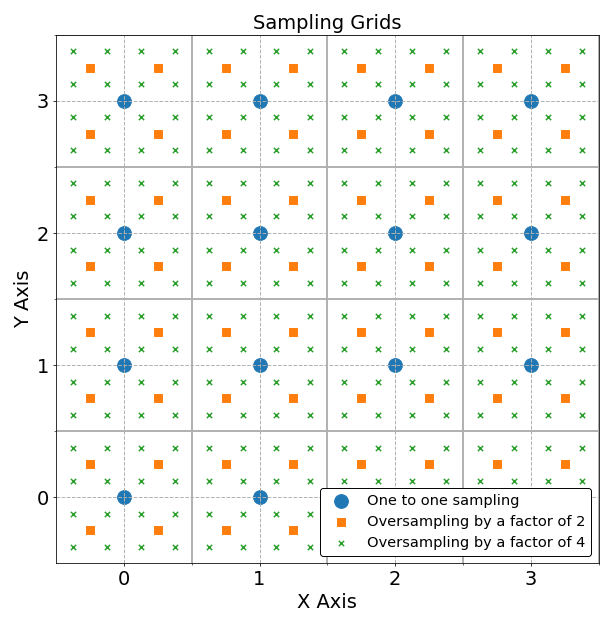}
\centering
\caption{\label{fig:sampling_grids} 
This plot shows three sampling grids; the blue circles represent one-to-one sampling, the orange squares represent an oversampling factor of two and the green  ``x" marks represent an oversampling factor of four. The boundaries of a pixel are repented by the solid lines and the pixel value is calculated by finding the mean of the sampling points. For clarity, only points of the same oversampling factor are used (one-to-one points are not sampled if oversampling factor is two, etc.).}
\end{figure}

One of the advantages of using PetroFit's \texttt{PSFModel} is its ability to sample models onto model images. For images that suffer from poor resolution, \texttt{PSFModel} samples the intensity model at sub-pixel levels and computes area-weighed sums (mean block-reduce) to approximate the integrated flux of each pixel (see Equation \ref{eq:oversampling}). \texttt{PSFModel} can oversample the entire model image or a specific pixel region of the image. An oversampling factor and region can be specified and passed as an argument when wrapping an Astropy model or during run-time by setting the \texttt{PSFModel.oversample} attribute. The following oversampling schemes are currently available: 

\begin{description}[leftmargin=0pt]

\item [Disabling Oversampling] To disable oversampling or one-to-one sampling, the oversampling can be set to Python \texttt{None}.

\item [Oversample Entire Model Image] To oversample the entire image by an oversampling factor, the oversampling factor can be passed as a single integer value. 

\item [Oversample a Fixed Region] To oversample a fixed square region of finite size the center pixel, length of the square region and an oversampling factor can be specified as a Python tuple.

\item [Oversample a Moving Region] While the model is being fit, the center of the model is likely to move around. To account for this, the names of model parameters that define the center of the box can be specified such that a square region of finite size ``follows" the model center and oversamples that region. 
\end{description}

For a grid that is oversampled by a factor of $N$ and mean block-reduced to match the data resolution, each pixel value ($F$) can be given with the following expression (see Figure \ref{fig:sampling_grids}):

\begin{equation}\label{eq:oversampling}
    F(i, j) \approx \frac{1}{N^2} \sum_{a=0}^{N-1} \sum_{b=0}^{N-1} I\left(x(i, a), y(j, b)\right)
\end{equation} 

Where:

$N$ is the oversampling factor.

$i$ and $j$ are the pixel indices.

$I$ is the intensity profile as a function of $x$ and $y$.

\vspace{2mm}

$x(i, a) = i + \frac{(1 + 2 a - N)}{2 N}$

\vspace{2mm}

$y(j, b) = j + \frac{(1 + 2 b - N)}{2 N}$

\vspace{2mm}

An issue that may come with sampling models onto images is making sure the peak value of the intensity profile is sampled into a pixel (in case it is between the sampling points). For concentrated profiles, this is especially important. For this reason, PetroFit allows for the center of the profile or model to be sampled by the sub-pixel nearest to it (replaces the sub-pixel value before reduction to data resolution).

\subsubsection{Levenberg-Marquardt Fitting}

PetroFit uses Astropy's \texttt{modeling} fitting algorithms to fit models such as \texttt{PSFModel}. A Levenberg-Marquardt least-squares fitting algorithm that utilizes a damping parameter to interpolate between the Gauss–Newton algorithm and the method of gradient descent \citep{levenberg1944AMF, marquardt1963AnAF} is used for nonlinear two dimensional models. Astropy implements this algorithm using SciPy's \texttt{optimize.leastsq} function, which is a wrapper around MINPACK’s \texttt{lmdif} and \texttt{lmder} algorithms \citep{minpack}.

\section{Demonstration of the Software}\label{sec:testofsoftware}

In this Section, we showcase capabilities outlined in the prior Sections by testing the software on simulated and real data sets. To test the accuracy of PetroFit's parameter estimates, we perform photometry on synthetic images of galaxies (generated by sampling Sérsic models onto pixel grids) and predict their Sérsic parameters. We also test PetroFit's fitting capability by modeling the light profiles of real galaxies in optical and near-infrared images taken by the Hubble Space Telescope and the Sloan Digital Sky Survey (SDSS).   

\subsection{Synthetic Sérsic Images}

We first demonstrate photometric measurements made by PetroFit in ideal conditions by simulating images of Sérsic profiles with no noise. We simulate Sérsic profiles described in Section \ref{sersicindex}, namely Gaussian ($n = 0.5$), exponential ($n = 1$), and de Vaucouleurs' ($n = 4$). To simulate the images, we defined Astropy \texttt{Sersic2D} models and placed the center of the models at the center of the output image. We set the image dimensions to $3500$ pixels, a large image size in order to include the flux without encountering any image edge effects in the measurement. For each simulated profile the parameters of the Sérsic model were set to $(I_e=1,\ r_e=50,\ x_0=1750,\ y_0=1750,\ ellip=0,\ \theta=0)$. We used \texttt{PSFModel} to sample the model onto a pixel grid (note that \texttt{PSFModel} is independent of the photometric measurement machinery), and applied an oversampling factor of $10$ to the center $100\times100$ pixels for all the model images. 

 Photometric measurements were then performed on the simulated profiles. First, the synthetic images were segmented using the \texttt{make\_catalog} catalog function. The curve of growth for each image was constructed using \texttt{source\_photometry} without background subtraction. A Petrosian profile was constructed using the curve of growth. The total flux radius was defined using the Petrosian profile and the default $\epsilon$ value of 2. We refer to the radii and concentrations derived using the default $\epsilon$ value as ``uncorrected.'' We also use a correction grid to generate an improved estimate of $\epsilon$ using the \texttt{PetrosianCorrection} class. 

Tables \ref{table:n05}, \ref{table:n1} and \ref{table:n4} show the resulting measurements. The ``Sersic" column represents analytical values derived from equations in Section \ref{sec:intro}, the ``Uncorrected" column shows measurements derived using $\epsilon = 2$, and the ``Corrected" column shows values derived using improved estimates of $\epsilon$. The values in parentheses show the ratios of measurements to their corresponding analytical values. As can be seen in these results, the corrected photometric values provide reasonably accurate measurements for these ideal cases.  

\begin{table}

\centering
\textbf{Synthetic Sérsic Image Measurements}

\vspace{1mm}

\caption{Gaussian Profile $n=0.5$ \label{table:n05}}
\begin{filecontents*}{simulated_2d_sersic_params_n0.5.csv}
Values,Sersic,Uncorrected,Corrected
eta,0.2,0.2 (1.0),0.2 (1.0)
epsilon,1.3,2.0 (1.54),1.31 (1.01)
r\_petrosian,98.77,98.77 (1.0),98.77 (1.0)
r\_half\_light,50.0,50.01 (1.0),49.76 (1.0)
r\_total\_flux,128.88,197.54 (1.53),128.98 (1.0)
total\_flux,22661.8,22660.98 (1.0),22436.45 (0.99)
r\_20,28.21,28.26 (1.0),28.26 (1.0)
r\_80,75.26,76.27 (1.01),75.27 (1.0)
c2080,2.13,2.16 (1.01),2.13 (1.0)
\end{filecontents*}
\csvautotabular{simulated_2d_sersic_params_n0.5.csv}

\vspace{2mm}

\caption{Exponential Profile $n=1$ \label{table:n1}}
\begin{filecontents*}{simulated_2d_sersic_params_n1.0.csv}
Values,Sersic,Uncorrected,Corrected
eta,0.2,0.2 (1.0),0.2 (1.0)
epsilon,1.82,2.0 (1.1),1.8 (0.99)
r\_petrosian,108.77,108.77 (1.0),108.77 (1.0)
r\_half\_light,50.0,49.76 (1.0),49.51 (0.99)
r\_total\_flux,197.76,217.54 (1.1),195.66 (0.99)
total\_flux,29871.24,29703.68 (0.99),29553.21 (0.99)
r\_20,24.39,24.5 (1.0),24.5 (1.0)
r\_80,87.64,88.27 (1.01),87.52 (1.0)
c2080,2.78,2.78 (1.0),2.76 (0.99)
\end{filecontents*}
\csvautotabular{simulated_2d_sersic_params_n1.0.csv}

\vspace{2mm}

\caption{De Vaucouleurs' Profile $n=4$  \label{table:n4}}
\begin{filecontents*}{simulated_2d_sersic_params_n4.0.csv}
Values,Sersic,Uncorrected,Corrected
eta,0.2,0.2 (1.0),0.2 (1.0)
epsilon,10.21,2.0 (0.2),9.37 (0.92)
r\_petrosian,92.77,92.77 (1.0),92.77 (1.0)
r\_half\_light,50.0,36.76 (0.74),49.01 (0.98)
r\_total\_flux,947.18,185.54 (0.2),868.95 (0.92)
total\_flux,56663.08,47160.67 (0.83),55967.71 (0.99)
r\_20,13.81,11.5 (0.83),13.75 (1.0)
r\_80,152.71,91.27 (0.6),151.53 (0.99)
c2080,5.22,4.5 (0.86),5.21 (1.0)
\end{filecontents*}
\csvautotabular{simulated_2d_sersic_params_n4.0.csv}

\vspace{1mm}

\end{table}

\subsection{Single Component Galaxy}\label{singlesection}

Real images of galaxies contain background noise and are smeared by a PSF. In this subsection, we demonstrate how the different tools in PetroFit can be used together to do a comprehensive analysis of galaxy light profiles. We first perform photometric measurements and estimate the Sérsic parameters for the galaxy. Using the estimated Sérsic parameters as an initial guess, we construct a \texttt{PSFModel} and fit the galaxy image. 

\begin{figure*}
\includegraphics[width=\textwidth]{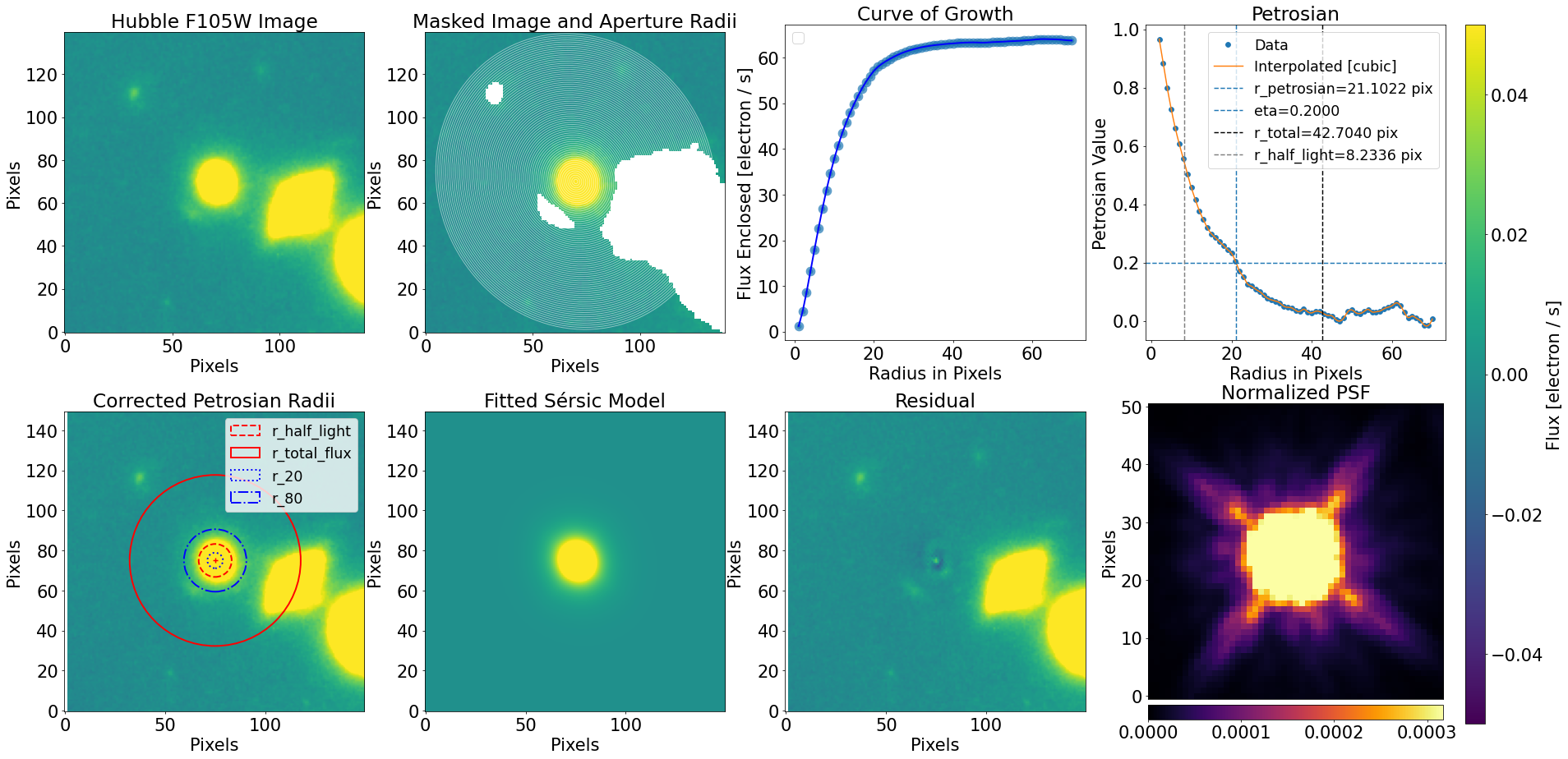}

\centering
\caption{\label{fig:singlegalaxy} 
The inputs, steps, and results of fitting a single component galaxy are shown in this figure. The top row shows, going from left to right: (a) The background-subtracted input F105W image of the target, (b) The masked image used for photometric measurements (also over-plots the apertures as white ellipses), (c) The curve of growth from photometric measurements as a function of aperture radius, (d) The Petrosian profile derived from the photometric measurements. The bottom row shows, going from left to right: (e) An over-plot of Petrosian radii to show their extent, (f) The fitted Sérsic model sampled onto an image (PSF convolved), (g) Residual produced by subtracting the model image from the original data, (h) The normalized PSF (cutout of a bright star) that was used for fitting.}
\end{figure*}

The data we use in this Section is a cutout of a group of bright galaxies in Abell 2744. The original data has a resolution of 60 milliarcseconds per pixel and was acquired by the Hubble Frontier Fields \citep{Lotz2017ApJ} team via the WFC3 instrument (using the F105W filter). This data-set can be directly \href{https://archive.stsci.edu/pub/hlsp/frontier/abell2744/images/hst/v1.0/}{downloaded from the Mikulski Archive for Space Telescopes}. The galaxy we selected for this demonstration is an elliptical galaxy $(\alpha=3.596248^{\circ}, \delta=-30.388517^{\circ})$ that can be well described by a single component Sérsic profile (see the center of the first panel in Figure \ref{fig:singlegalaxy}). A $150 \times 150$ pixel cutout of the target was made for this demonstration and its background was subtracted by fitting a 2D plane (see Section \ref{backgroundNoiseSubtraction}) to a 3 sigma clipped version of the cutout. We refer to this cutout as the galaxy image to avoid confusion with other cutouts.

\subsubsection{Photometry of Single Component Galaxy}\label{singlephoto}

 Since the target galaxy has nearby neighbors, we first made a catalog and segmentation map using the \texttt{make\_catalog} function. The threshold was computed using Astropy's \texttt{sigma\_clipped\_stats} with a $2$ sigma cutoff (this resulted in the best deblending). A smoothing kernel of size $3 \times 3$ and $fwhm=3$ was used for the segmentation and deblending steps. The $npixels$ was set to $4$ pixels and the deblending contrast was set to $0$. 

We constructed a curve of growth (COG) using the \texttt{source\_photometry} function. To construct the COG we used an array of $70$ aperture radii up to $70$ pixels ($1$ aperture per pixel). We set the photometry cutout size (not to be confused with the galaxy image) to twice the maximum pixel.  \texttt{source\_photometry} was used to make a measurement cutout of the specified size, mask nearby sources and make photometric measurements using the defined apertures.  Once the COG was constructed using the photometric measurements, the galaxy's Petrosian profile was computed and a corrected $\epsilon$ value was estimated using a correction grid as described in Section \ref{sec:software}.

\subsubsection{Single Component Fitting}\label{singlefitting}

We defined a \texttt{Sersic2D} model and wrapped it with \texttt{PSFModel}. We set the oversampling box ($20$ pixels on each side) to follow the center of the model and oversample by a factor of $5$. PSF rotation was disabled and \texttt{psf\_pa} was set to $0$.  Using the calculated photometric quantities, we set initial estimates for the model of the object based on the following measurements:

\begin{description}[leftmargin=0pt]

\item [\texttt{amplitude}] The amplitude was estimated by fitting an elliptical isophot at \texttt{r\_eff} using \href{https://photutils.readthedocs.io/en/stable/isophote.html}{\texttt{photutils.isophot}}.

\item [\texttt{r\_eff}] The effective radius was set using the corrected Petrosian half-light radius.
\item [\texttt{n}] The Sérsic index was estimated using the correction grid through a reverse lookup of the Sérsic index that produces the uncorrected Petrosian radius and concentration ($C2080$).
\item [\texttt{x\_0 \& y\_0}] The center was set to the pixel with maximum flux value.
\item [\texttt{ellip} and \texttt{theta}] Ellipticity and the rotation angle were set using the Photutils measurement during the segmentation and deblending steps.
\item[\texttt{PSFModel}]  A cutout of a nearby, bright star was used to model the PSF.  The image cutout was  normalized and had a size of $51 \times 51$ pixels. 
\end{description}

The Astropy \texttt{LevMarLSQFitter} fitter was then used to fit the model. The fitter running on an Intel Core i9-10900K Processor (20MB Cache, 3.7GHz) completed the fit under $0.1$ seconds. Figure \ref{fig:singlegalaxy} shows the resulting fit and associated residual.

We compare the initial guesses obtained from photometric and Petrosian measurements before the fitting to the parameters estimated by the fitter in Table \ref{table:singleparamcomp}.

\begin{table}
\caption{Table of Estimated and Fitted Parameters}\label{table:singleparamcomp}
\begin{tabular}{ | l | c | c | }
\hline
Parameter & Initial Guess & Fitted \\
\hline
amplitude & 0.083 & 0.087 \\
r\_eff & 8.234 & 7.737 \\
n & 1.388 & 1.621 \\
x\_0 & 75.000 & 75.906 \\
y\_0 & 75.000 & 75.408 \\
ellip & 0.084 & 0.109 \\
theta & -1.130 & -1.136 \\
psf\_pa & 0.000 & 0.000 \\
\hline
\end{tabular}
\end{table}

Using \texttt{sersic\_enclosed} we compute the analytical total Sérsic flux of the model at infinity and compare it to the photometrically derived Petrosian total flux of the galaxy in the image. The total Sérsic flux is $21.684\ mags$ while the corrected Petrosian total flux of the galaxy image was measured to be $21.7593\pm 0.0026\ mags$. The corrected Petrosian total flux of the model was measured to be $21.75899\ mags$, which is $93\%$ of the total Sérsic flux (see Section \ref{petrosianepsilon}). 

\subsection{Multi-Component Galaxy}\label{multicompgalaxy}

We further demonstrate PetroFit's fitting capabilities by fitting the light profile of Messier 91 (M91), a barred galaxy with multiple Sérsic components. Unlike single component galaxies, multi-component light profiles require careful and complex decompositions. To achieve this, we used multi-band SDSS images ($1000 \times 1000$ pixels covering $20$ arcminutes) centered on M91 $(\alpha=188.860221^{\circ}, \delta=14.496339^{\circ})$.  After subtracting the background by fitting a plane to a 3 sigma clipped version of the image, we used the i band to estimate the morphological parameters of the source as described in Section \ref{singlephoto}. We derived rough estimates for our initial guesses by following the steps outlined in Section \ref{singlefitting}. For fitting the galaxy, we zoomed into the galaxy by creating a cutout of ($300 \times 300$ pixels). Using the rough parameter guess we derived, we defined three Sérsic components, one for the disk and two oversampled components for the core and bar. Using a nearby star $(\alpha=188.740014^{\circ}, \delta=14.495436^{\circ})$ as a PSF, we found that the image had enough resolution such that the PSF was not needed for the galaxy decomposition; therefore we disabled that feature for all components. For the disk, we applied a sigma mask to mask out the core and bar of the galaxy. We then fit the disk using this masked image to better constrain our initial guess. The core and bar components were oversampled by a factor of $5$. For the core, we masked out the disk and bar using an elliptical mask that was visually parameterized. Lastly, the bar was fit after masking out the core and disk using a mask that was also visually parameterized. These three preliminary fits bring the components closer to the final fit. We combined the three components into a compound model and wrapped it in a \texttt{PSFModel} model with an oversampling factor of $5$. We then fit the entire galaxy cutout, which resulted in the final i band fit. 

\begin{figure*}[t]
\includegraphics[width=\textwidth]{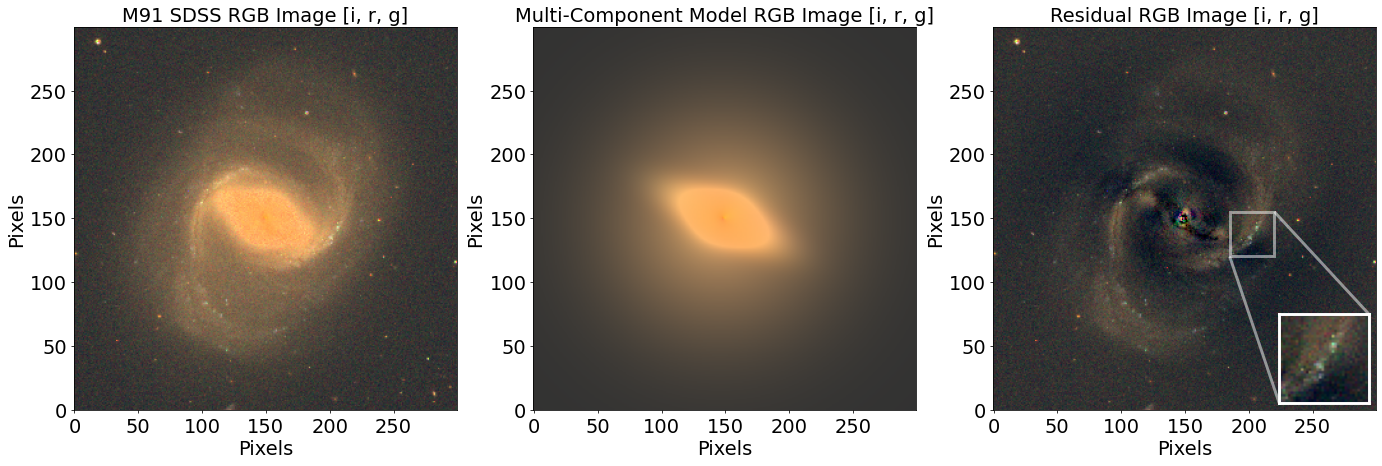}

\centering
\caption{\label{fig:multicomp} 
These panels show Lupton-RGB ([i, r, g] band) images of Messier 91 (M91), a corresponding multi-component Sérsic fit, and the resulting residual. Three components were used to fit this light profile, one for the disk, bar, and central core. The images were made using a stretch factor of $0.9$ and $Q=2$. The first panel shows the SDSS M91 cutout that was used for fitting, the middle panel shows the fitted model image, and the last panel shows the residual. The spiral arm, as well as blue clusters (zoomed-in white box), can also be seen in the residual image.     
}
\end{figure*}

To perform a full analysis of the light profile, we also fit the galaxy in g, r, and z bands. For each of these bands, the fit from i band was used as the initial guess of the compound model parameters. This allowed us to skip the photometry, Petrosian, and the preliminary fitting steps for these bands. After fitting the models, we combine the i, r and, g bands into an RBG image to illustrate the galaxy, fitted model, and residual (see Figure \ref{fig:multicomp}). The resulting residual image clearly shows the spiral arms of M91, as well as blue clusters embedded in it. The bar component over-subtracts the profile, mostly because the radial profile used was elliptical as opposed to ``boxy" and because actual galaxy bars quickly truncate at their edges. We plan on introducing various radial profiles, including ``boxy" profiles, in the future versions of PetroFit (see Section \ref{sec:conc} and Figure \ref{fig:custommodes}). The final fits running on an Intel Core i9-10900K Processor (20MB Cache, 3.7GHz) completed the fit under $2.5$ minutes per band. For the core of the galaxy, each band had a Sérsic index of approximately $4$ while the disk and bar components had an index of $0.5$.  

\subsection{Simultaneously Fitting Overlapping Galaxies}\label{overlappinggalaxies}
 
\begin{figure*}

\centering

\includegraphics[width=\textwidth]{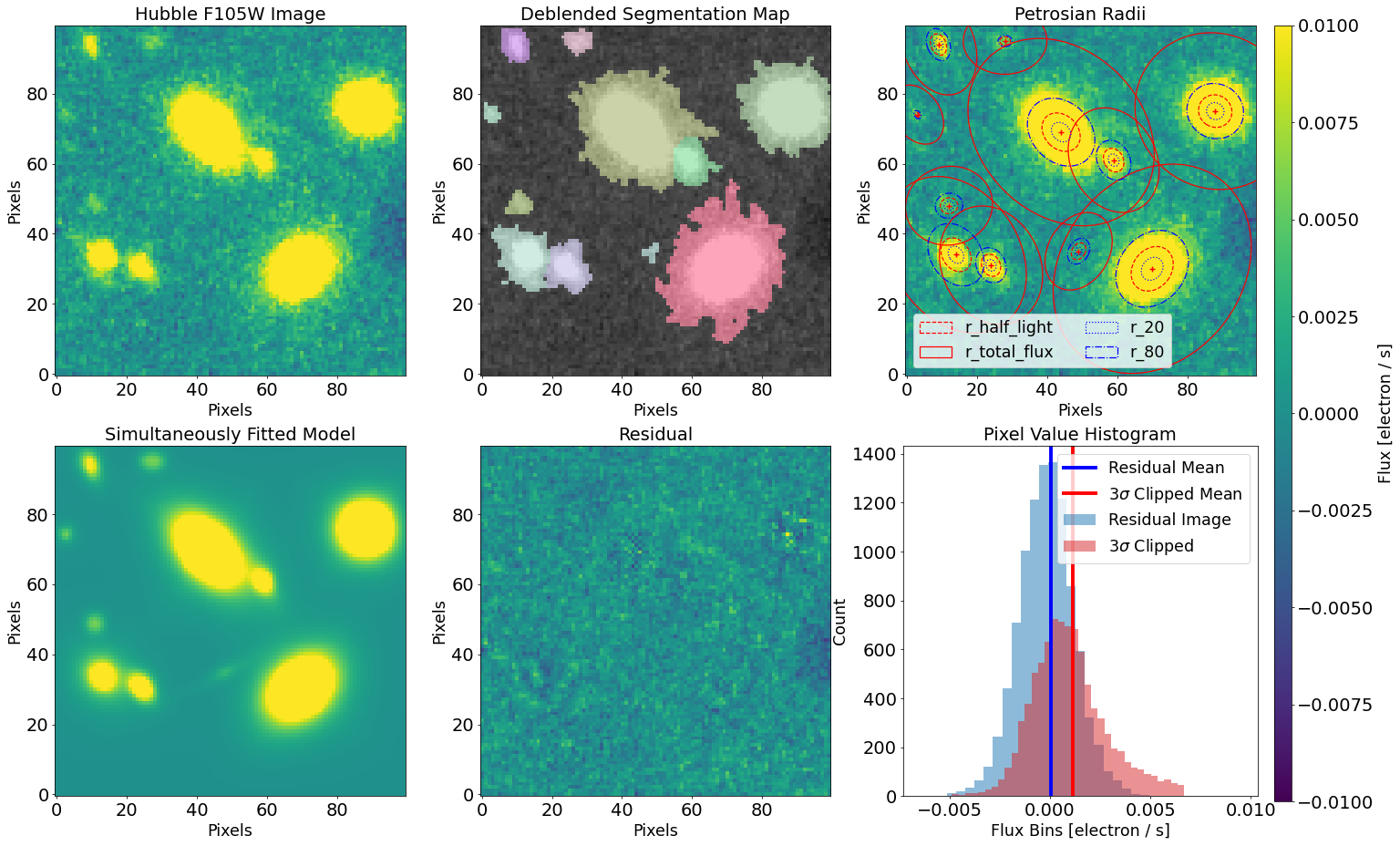}

\caption{\label{fig:multiegalaxy} 
This plot shows some of the inputs, steps, and results of fitting a  multi-component model. The top row shows, going from left to right: (a) The input F105W image of the group of galaxies, (b) Deblended segmentation map with color-coding for pixels that belong to the same source, (c) An over-plot of Petrosian radii of each source. The bottom row shows, going from left to right: (d) The fitted compound model, with $11$ Sérsic components, sampled onto an image (PSF convolved), (e) Residual image produced by subtracting the model image from the original data, (f) A histogram of pixel values in the residual image and a $3\sigma$ clipped version of the original image. Note that the residual image histogram takes on a Gaussian form centered closer to $0$ than the original image.}
\end{figure*}

A key capability of PetroFit is its ability to simultaneously fit and measure the photometry of multiple sources. Images of galaxies with nearby neighbors pose a challenge because their light profiles overlap. This is especially true in cluster environments. In this Section, we demonstrate how PetroFit can be used to construct photometry catalogs of overlapping sources and fit compound models. The group of galaxies we selected for this measurement are a part of the Hubble Frontier Fields data discussed in Section \ref{singlephoto}. The group of galaxies located at $(\alpha = 3.579222^{\circ},\ \delta = -30.380186^{\circ})$ are faint and isolated from bright sources.  The image is background-subtracted using a 2D plane model as described in Section \ref{backgroundNoiseSubtraction}. We first make photometric and size measurements to inform our initial guesses. We construct a catalog using \texttt{make\_catalog} with the following settings: the detection threshold was set to the standard deviation computed using \texttt{sigma\_clipped\_stats} ($3 \sigma$ clipping), a smoothing kernel size of $3$ pixels, smoothing kernel fwhm of $3$, minimum of $16$ connected pixels above the threshold, $0$ contrast and deblending enabled. This results in the identification of $11$ galaxies in the image. Once a Photutils catalog is generated, we loop through the catalog and calculate the Petrosian profile of each source. From the Petrosian profiles, we compute the sizes of each source.

For the fitting step, we construct a \texttt{Sersic2D} model and make initial guesses for each object as described in Section \ref{singlefitting}, except we do not apply Petrosian corrections and assume a Sérsic index of $1$. This is because the sources are likely to have a low concentration and to demonstrate that the initial guesses can be rough estimates. We sum the \texttt{Sersic2D} models together to create a single compound model. The compound model will have parameters corresponding to each sub-model's parameters, resulting in $77$ parameters in total ($11$ sub-models times $7$ parameters for each \texttt{Sersic2D} sub-model). Fitting such a large parameter space is possible because the initial guesses help the fitter converge to a solution. We wrap the compound model in a \texttt{PSFModel} and set it to oversample the entire image by a factor of $4$. We use the same star cutout as Section \ref{singlefitting} as the convolving PSF. The resulting \texttt{PSFModel} is fit using a Levenberg-Marquard fitter. The fitter, running on an Intel Core i9-10900K Processor (20MB Cache, 3.7GHz-5.3GHz), was able to simultaneously fit the $11$ component model under $4.5$ minutes. Figure \ref{fig:multiegalaxy} shows the results. 

\subsection{Large Scale Photometry Catalogs}

\begin{figure*}
\includegraphics[width=\textwidth]{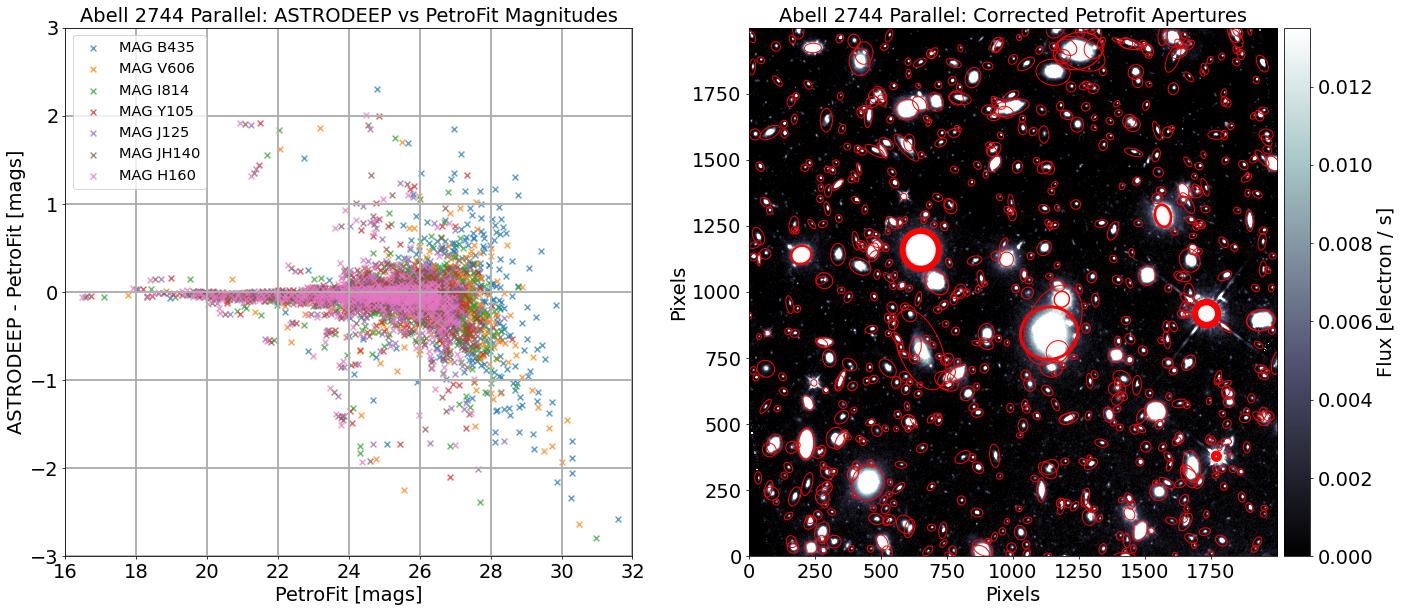}
\centering
\caption{\label{fig:abell_2744_parallel}
These plots show, going from left to right: (a) Multi filter absolute AB magnitude comparison between PetroFit aperture photometry and the ASTRODEEP catalog. (b) PetroFit Petrosian apertures plotted over the F105W image (detection image). 
}
\end{figure*}

In this Section, we perform large-scale multi-band aperture photometry on the Hubble Frontier Fields Abell 2744 parallel field images and compare it to the ASTRODEEP catalogs \citep{astrodeep2016A&A}. To catalog and measure the sizes of sources in the image, we  selected the F105W filter as our detection image. We estimated the image background by taking a patch in the image with no sources and measuring its standard deviation, which we used to define a segmentation threshold of $3 \sigma$. We segmented and deblended the image using a kernel size of 3 pixels, $fwhm = 3$, and a minimum source size of 16 pixels. For the deblending step, we set the contrast to $1/100$. 

Since we planned to use Petrosian radii as apertures for our catalog, we measured the curve of growth and Petrosian radii of each source in the segmentation map we generated. Petrosian corrections were applied to sources with uncorrected half-light radii greater than $10$ pixels ($0.6$ arcsec). If the Petrosian profile could not be measured for a source, we used two times the bounding box of the sources' segmentation as an estimate of the aperture radius. Using the Petrosian total flux radii as aperture sizes, we made photometric measurements of each source in the F435W, F606W, F814W, F105W, F125W, F140W, and F160W filters. This was conveniently possible (without having to segment each image) because images are pixel aligned and have the same pixel scales ($60$ mas/pixel). This allowed us to apply the same apertures to all filters. We applied local background subtractions by fitting a plane to one sigma pixels in cutouts of targets. Figure \ref{fig:abell_2744_parallel} shows the results from the multi-aperture measurements. The first plot shows a magnitude comparison between the PetroFit measurements and the corresponding ASTRODEEP catalog.

\section{Comparison to other Packages}\label{sec:otherpackages}

In this Section, we make feature comparisons between PetroFit and similar packages for measuring galaxy properties. We also provide comparative morphology and fitting tests in Appendix \ref{apdx:comparison}. 

Statmorph \citep{rodriguezGomez2019MNRAS} is a package for calculating non-parametric morphological diagnostics of galaxy images and fitting single component 2D Sérsic profiles. Statmorph uses radial profiles, including the Petrosian profile, to compute radii of interest and concentrations. PetroFit includes features, such as correction grids, to account for PSF smearing during parameter estimations. Statmorph's offers a single component Sérsic fitting that is compatible with Astropy while PetroFit can fit multiple components of any 2D model variety. PetroFit fitting code also provides oversampling schemes for fitting images of galaxies with poor angular resolutions or profiles with high concentrations. At the time of writing, statmorph offers a variety of morphological parameters, such as Gini-M20 statistics, that are currently not implemented in PetroFit. Optimizing PetroFit's functions and including some of these features or investigating potential synergies between Statmorph and PetroFit is work for future versions of PetroFit. Though either package can be used for many overlapping science cases, PetroFit was built specifically for making precision angular size measurements and offers tools to fit complex light profiles (See Sections \ref{multicompgalaxy} and \ref{overlappinggalaxies}). 

GALFIT \citep{peng2010} has and continues to serve the community with its robust galaxy light profile fitting code.  This package uses modeling techniques similar to PetroFit but also offers azimuthal shape functions that can be used to define shapes beyond elliptical distributions. Similar types of capabilities are planned for implementation in future iterations of PetroFit (see Section \ref{sec:conc} and Figure \ref{fig:custommodes}). PetroFit provides non-parametric estimates of galaxy properties beyond model fitting. It also provides a python-based package that is compatible with the Astropy ecosystem, allowing users to easily interact with fitted compound models. PetroFit offers tools to make initial guesses of model parameters. Lastly, PetroFit allows for custom models or parameters to be defined on the fly by virtue of Astropy's custom modeling.  

\section{Conclusion} \label{sec:conc}
 
\begin{figure*}[t]
\includegraphics[width=\textwidth]{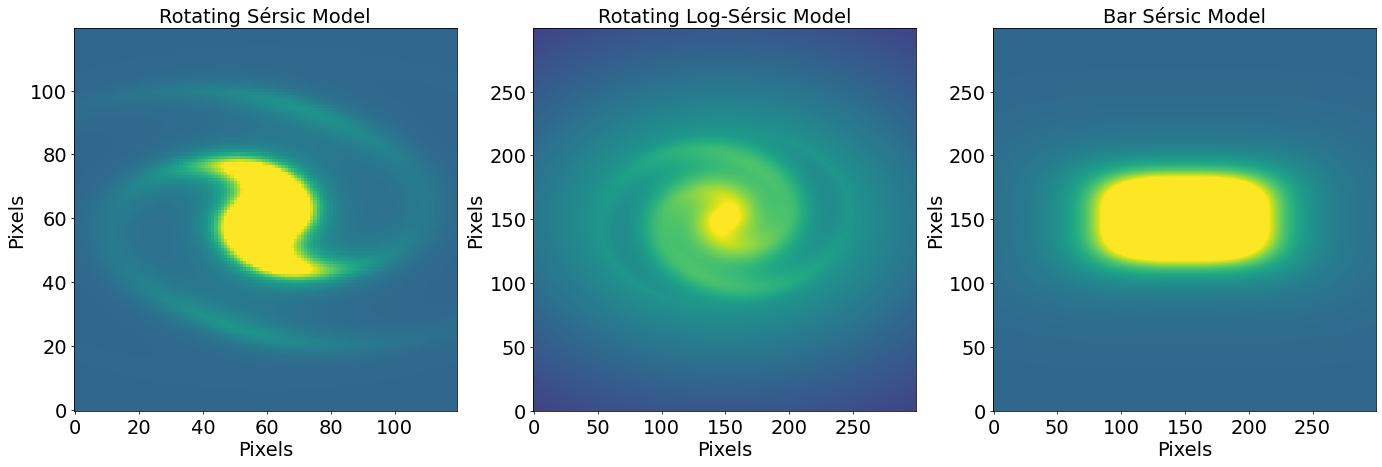}

\centering
\caption{\label{fig:custommodes} 
Examples of azimuthal shape functions implemented as custom 2D Astropy (fittable) models. Inspired by PHIR (Peng-Ho-Impey-Rix) models \citep{peng2010}, the first and middle panel show Sérsic models that apply simple rotations to the profile as a function of radius (i.e $\theta(r[x, y], m, r_{in})$, where m is the number of full rotations and $r_{in}$ is the rotation cutoff). The last panel shows a barred Sérsic profile that was generated using a generalized ellipse.}
\end{figure*}

In this paper, we introduce PetroFit, an open-source Python package for fitting galaxy light profiles and measuring Petrosian properties. PetroFit offers a wide range of tools for making accurate photometric measurements, constructing curves of growth, and computing Petrosian profiles. Computed Petrosian profiles can be used to estimate the projected radii of galaxies as well as their concentrations.  PetroFit includes a 2D image fitting tool-set that is built using the modeling package in Astropy. We demonstrate capabilities, limitations, and potential use cases for the software by analyzing images of simulated and real galaxy light profiles. 

The version of PetroFit described in this paper represents an initial step towards our goal of a robust package for making galaxy size measurements. We plan on continuously improving the software and hope to gain useful feedback from the Astronomical community. In the next iterations, we hope to add features such as better error estimations in fitted models, Astropy unit support for Petrosian measurements, and advanced machine learning-based parameter estimates (an experimental decision tree regressor is currently available for Petrosian corrections). In future versions of PetroFit, we hope to allow for corrections to be computed using $\epsilon_{80}$ and $P_{0502}$. We also hope to officially implement azimuthal shape functions defined by \cite{peng2010}, which we refer to as PHIR (Peng-Ho-Impey-Rix) models, which use Fourier and bending modes to create complex profiles (See Figure \ref{fig:custommodes} for examples of simple azimuthal Astropy models). We aim to make PetroFit more compatible with similar packages for the purposes of interoperability and testing. As part of this effort, we also hope to contribute to PhotUtils and Astropy features that have been developed here that may be useful beyond the goals of PetroFit. 

Due to the ability to accurately measure galaxy photometry and sizes systematically, PetroFit can support a wide range of galaxy evolution studies including classification of galaxies, assessing changes in the size of galaxies, and measurements in the luminosity function of galaxies. The flexible, object-oriented design of the code makes it ideal for use in the next generation of high-resolution galaxy surveys that will be produced by JWST and the Roman Space Telescopes. 

We openly invite and strongly encourage all interested parties to leave feedback, bug reports, and feature requests by opening tickets in the \href{https://github.com/PetroFit/petrofit}{\texttt{PetroFit/petrofit}} GitHub repository. We hope this package is as useful for the general astronomical community as it has been for the authors.

\newpage
\acknowledgments

 We thank our collaborators Dr. Gregory D. Wirth and Dr. D.J. Pisano for their close guidance and support throughout this project. We would like to thank Alan Kahn (Department of Physics, Columbia University) for useful conversations about machine learning and its potential applications. We would like to thank the developers of PhotUtils as much of this work is built on top of the tools they provided.
 
 This work was made possible by NASA grant AR 15058.  We give thanks to the Space Telescope Science Institute and the Technical Staff Research Committee (TSRC), a group that facilitates matching research projects with interested technical staff.  This research is based on observations made with the NASA/ESA Hubble Space Telescope obtained from the Space Telescope Science Institute, which is operated by the Association of Universities for Research in Astronomy, Inc., under NASA contract NAS 5–26555. 
 
 Funding for the SDSS and SDSS-II has been provided by the Alfred P. Sloan Foundation, the Participating Institutions, the National Science Foundation, the U.S. Department of Energy, the National Aeronautics and Space Administration, the Japanese Monbukagakusho, the Max Planck Society, and the Higher Education Funding Council for England. The SDSS Web Site is \url{http://www.sdss.org/}.

The SDSS is managed by the Astrophysical Research Consortium for the Participating Institutions. The Participating Institutions are the American Museum of Natural History, Astrophysical Institute Potsdam, University of Basel, University of Cambridge, Case Western Reserve University, University of Chicago, Drexel University, Fermilab, the Institute for Advanced Study, the Japan Participation Group, Johns Hopkins University, the Joint Institute for Nuclear Astrophysics, the Kavli Institute for Particle Astrophysics and Cosmology, the Korean Scientist Group, the Chinese Academy of Sciences (LAMOST), Los Alamos National Laboratory, the Max-Planck-Institute for Astronomy (MPIA), the Max-Planck-Institute for Astrophysics (MPA), New Mexico State University, Ohio State University, University of Pittsburgh, University of Portsmouth, Princeton University, the United States Naval Observatory, and the University of Washington.

 \software{
PetroFit \citep{petrofit}, 
Astropy \citep{astropy13}, 
Photutils \citep{bradley2020},
SciPy \citep{2020SciPy-NMeth}, 
Numpy \citep{2020NumPy-Array},
MatPlotLib \citep{Matplotlib},
Statmorph \citep{rodriguezGomez2019MNRAS},
GALFIT \citep{peng2010}
}

%%%%%%%%%%%%%%%%%%%%%%%%%%%%%%%%%%%%%%
%%%%%%%%%%%%%%%%%%%%%%%%%%%%%%%%%%%%%%
%%%%%%%%%%%  Appendix %%%%%%%%%%%%%%%%
%%%%%%%%%%%%%%%%%%%%%%%%%%%%%%%%%%%%%%
%%%%%%%%%%%%%%%%%%%%%%%%%%%%%%%%%%%%%%

\clearpage

\appendix

\section{List of Models}\label{apdx:listofmodels}

\subsection{Astropy}
\begin{center}

\href{https://docs.astropy.org/en/v4.2/modeling/predef_models2D.html}{All Astropy (\texttt{v4.2}) 2D Models}

\end{center}

\begin{center}
\begin{tabular}{  |p{4cm}|p{12cm}|  }
\hline
Model Name & Description \\
\hline
    \texttt{AiryDisk2D}  & Airy Disk.\\
    \texttt{Box2D} &  2D Box profile.\\
    \texttt{Disk2D} &  A Disk profile with a constant amplitude.\\
    \texttt{TrapezoidDisk2D} &  Disk with a slope.\\
    \texttt{Ellipse2D} & Ellipse with major and minor axis and rotation angle.\\
    \texttt{Gaussian2D} &  2D Gaussian (can be elliptical) distribution with a rotation angle.\\
    \texttt{Moffat2D} &  Moffat profile (can be elliptical) with alpha (power index) and gamma (core width).\\
    \texttt{RickerWavelet2D} &  Symmetric Ricker Wavelet function with the specified sigma.\\
    \texttt{Sersic2D} &  Sérsic profile with an effective half-light radius, rotation, and Sérsic index.\\
    \texttt{Ring2D} & A 2D Ring with inner and outer radii.\\
    \texttt{Const2D} & A 2D constant (flat plane). \\
\hline
\end{tabular}
\end{center}

\subsection{PetroFit}
% \begin{center}
%     PetroFit Models
% \end{center}

\begin{center}
\begin{tabular}{ |p{4cm}|p{12cm}| }
\hline
Model Name & Description \\
\hline
    \texttt{Nuker2D}  & 2D Nuker profile\\
    \texttt{CoreSersic2D} &  2D Core Sérsic profile\\
    \texttt{sersic\_enclosed\_model} & Total flux enclosed within a radius (Sérsic profile) \\
    \texttt{petrosian\_model} & Petrosian profile (Sérsic profile) \\
\hline
\end{tabular}
\end{center}

\clearpage

\section{Approximations of Sérsic and Petrosian Quantities}\label{paramapproximationss}

In this Section, we discuss various methods of approximating Sérsic and Petrosian quantities. In particular, we look at the relationships between Sérsic indices, corrected and uncorrected concentrations. We also cover why correction grids are necessary despite the availability of the approximations discussed in this Appendix. 

\subsection{Analytical Relationships}

\subsubsection{Sérsic Indices and Concentrations}

For purely analytical Sérsic profiles that extend to infinity, the concentration can be found by solving for the radii where $L(\leq r_{i}) =  \frac {i} {100} L(\leq \infty)$ and  $L(\leq r_{o}) = \frac {o} {100} L(\leq \infty)$. Using Equations \ref{eq:fluxenclosed} and \ref{eq:totalsersicfluxenclosed},  along with $x(r) = b_n (r / r_e)^{1/n}$, concentration radii can be expressed as follows:

\begin{equation}
    r_i = r_e \left(\frac{\gamma^{-1} \left(2n, \Gamma(2 n) \frac {i} {100} \right)}{b_n}\right)^n
\end{equation}

\begin{equation}
    r_o = r_e \left(\frac{\gamma^{-1} \left(2n, \Gamma(2 n) \frac {o} {100} \right)}{b_n}\right)^n
\end{equation}

\begin{equation}
    \frac{r_{o}}{r_{i}} = \left(\frac{\gamma^{-1} \left(2n, \Gamma(2 n) \frac {o} {100} \right)}{\gamma^{-1} \left(2n, \Gamma(2 n) \frac {i} {100} \right)}\right)^n
\end{equation}

\begin{equation}
    C_{io} = 5 \cdot log\left(\frac{r_{o}}{r_{i}}\right) = 5 \cdot n \cdot log{\left(\frac{\gamma^{-1} \left(2n, \Gamma(2 n) \frac {o} {100} \right)}{\gamma^{-1} \left(2n, \Gamma(2 n) \frac {i} {100} \right)}\right)}
\end{equation}

\subsection{Approximations}

We approximate the following relationships by first generating a grid of quantities using ideal Sérsc and Petrosian profiles for $0.5 \leq n \leq 15$. We then make corrected and uncorrected Petrosian measurements. Note that  ``uncorrected" means that Petrosian radii and concentrations were computed using the default $\epsilon = 2$. For each value of $n$, we followed the steps below to generate the profiles and quantities used to derive relationships:

\begin{enumerate}
  \item Calculate Sérsic radii containing fluxes of interest, including $r_{99\_sersic}$ (which we define as the total flux radius), using \texttt{sersic\_enclosed\_inv}.
  
  \item Construct a radius list to sample analytical functions discussed in Section \ref{analyticalprofiles}. 
  
  \item Use the radius list and \texttt{sersic\_enclosed} to construct a curve of growth (flux array). 
  
  \item Use the radius list and \texttt{petrosian\_profile} to construct an ideal Petrosian profile. We over-sample the inner $r_{80}$ by a factor of $10000$ to produce accurate measurements for high concentration profiles. For radii larger than $r_{80}$, we sample at a maximum of $1000$ points (equally spaced). 
  
  \item Using $\eta = 0.2$ and the ideal Petrosian profile, find $r_{petro}$. 
  
  \item Using $\epsilon = 2$ and $r_{petro}$, compute the uncorrected $r_{total}$, $r_{50}$, $r_{20}$, $r_{80}$ and $C_{2080}$.
  
  \item Find the corrected $\epsilon$ value using $\epsilon_{correct}  = r_{99\_sersic} / r_{petro}$.
  
  \item Using $\epsilon_{correct}$ and $r_{petro}$, compute the corrected $r_{total}$, $r_{50}$, $r_{20}$, $r_{80}$ and $C_{2080}$.
  
\end{enumerate}

The approximations in the Subsections that follow were derived by fitting models using Astropy modeling. 

\subsubsection{Corrected vs Uncorrected Concentration Index}

\begin{figure}[h]

\centering

\includegraphics[width=8cm]{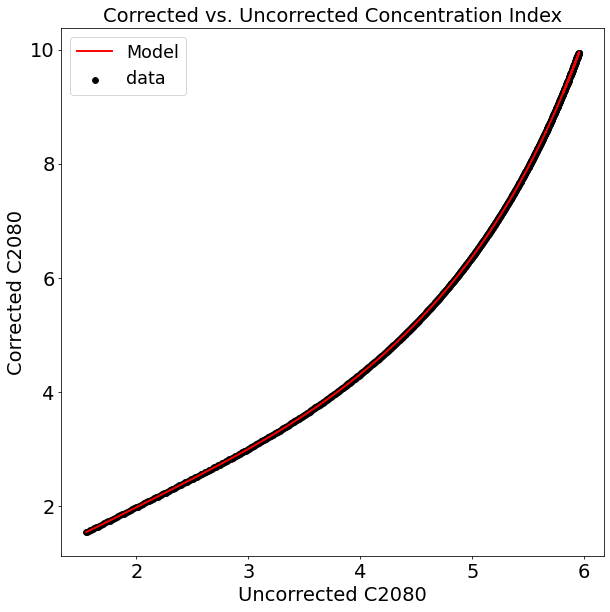}

\caption{
\label{corr_vs_uncorr_c2080}
Corrected $C_{2080}$ approximated from the uncorrected $C_{2080}$ for $0.5 \leq n \leq 15$. The black markers represent values obtained from measurements and the red line is the line of best fit.}
\end{figure}

Because uncorrected concentrations are derived from radii calculated using $\epsilon = 2$, errors in the radii estimates propagate to estimates of concentrations. This effect is especially severe for profiles with high concentrations. We estimate corrected $C_{2080}$ using the uncorrected $C_{2080}$  for $0.5 \leq n \leq 15$ using a sixth degree polynomial (see Figure \ref{corr_vs_uncorr_c2080}). For clarity we denote the corrected $C_{2080}$ with $C$ and uncorrected $C_{2080}$ with $U$:

\begin{equation}\label{eq:c_to_u}
\begin{array}{l}
    C(U) \approx 2.26194802 - 3.61130833 \cdot U + 3.8219758 \cdot U^2 - 1.6414660 \cdot U^3 \\
    + 0.38059409 \cdot U^4 -0.0450384 \cdot U^5 + 0.00221922 \cdot U^6
\end{array}
\end{equation}

\subsubsection{Sérsic Index vs. Concentration Index}\label{c_vs_n_approx}

\begin{figure}[h]

\centering

\includegraphics[width=16cm]{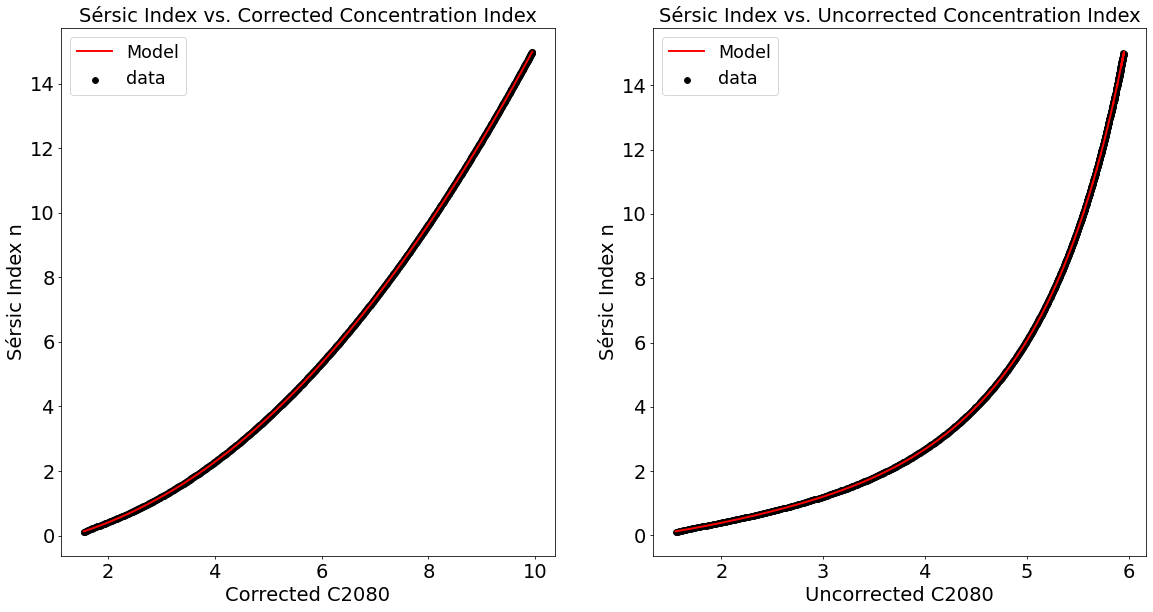}

\caption{
\label{c2080_vs_n}
Sérsic index $n$ estimated using the corrected and uncorrected $C_{2080}$ for $0.5 \leq n \leq 15$ using Equations \ref{eq:n_vs_c} and \ref{eq:n_vs_u} (red lines).}
\end{figure}

Since both the concentration index and Sérsic index are measures of concentration, they are correlated. We estimate the Sérsic index $n$ using the corrected $C_{2080}$  for $0.5 \leq n \leq 15$ using a fifth degree polynomial (see Figure \ref{c2080_vs_n}). The relationship between the uncorrected $C_{2080}$ and Sérsic index $n$ can be computed first computing the corrected $C_{2080}$  using Equation \ref{eq:c_to_u}. For clarity we denote the corrected $C_{2080}$ with $C$ and uncorrected $C_{2080}$ with $U$:

\begin{equation}
\begin{array}{l}\label{eq:n_vs_c}
    n(C) \approx -0.41844073 + 0.20487513 \cdot C + 0.08626531 \cdot C^2 \\
    + 0.0106707 \cdot C^3 - 0.00082523 \cdot C^4 + 0.00002486 \cdot C^5 
\end{array}
\end{equation}

\begin{equation}\label{eq:n_vs_u}
    n(U) \approx n(C(U))
\end{equation}

\subsubsection{Epsilon vs. Concentration Index and Sérsic Index} \label{epsilonvsc}

\begin{figure}[h]

\centering

\includegraphics[width=\linewidth]{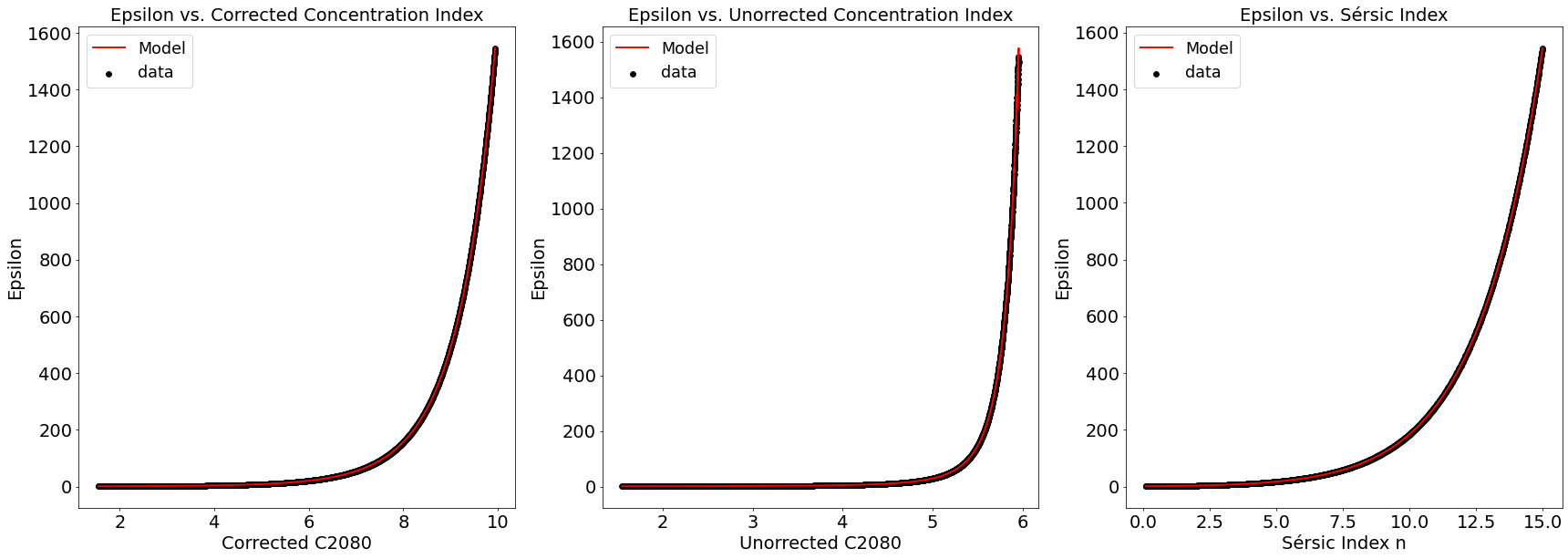}

\caption{
\label{epsilon_vs_n_and_c2080}
Epsilon ($\epsilon$) estimated using the Sérsic index $n$, corrected and uncorrected $C_{2080}$  for $0.5 \leq n \leq 15$ using Equation \ref{eq:n_vs_epsilon_approx} and \ref{eq:c_vs_epsilon_approx} (red lines)}
\end{figure}

 Epsilon ($\epsilon$) is the multiplying factor that is used to compute $r_{99}$. We estimate epsilon ($\epsilon$) using the Sérsic index $n$ using a combination of a fifth degree polynomial and an exponential (see Figure \ref{epsilon_vs_n_and_c2080}). We also use the formulations in Subsections \ref{c_vs_n_approx} to compute the relationship between $\epsilon$ and  $C_{2080}$. For clarity we denote the corrected $C_{2080}$ with $C$ and uncorrected $C_{2080}$ with $U$:
 
 \begin{equation}\label{eq:n_vs_epsilon_approx}
 \begin{array}{l}
    \epsilon(n) \approx -6.54870813 - 2.15040843 \cdot n - 0.28993623 \cdot n^2 - 0.04099376 \cdot n^3 \\
    - 0.00046837 \cdot n^4 - 0.00022305 \cdot n^5 + 7.48787292 \cdot exp\left\{ \frac{n}{2.6876055} \right\}
\end{array}
\end{equation}

 \begin{equation}\label{eq:c_vs_epsilon_approx}
 \begin{array}{l}
   \epsilon(C) \approx \epsilon( n(C) )\\
   \epsilon(U) \approx \epsilon( n(C(U)) )
   
\end{array}
\end{equation}

\subsubsection{Epsilon vs. Petrosian Concentration Index}\label{epsilonvsp}

\begin{figure}[h]
\centering

\includegraphics[width=8cm]{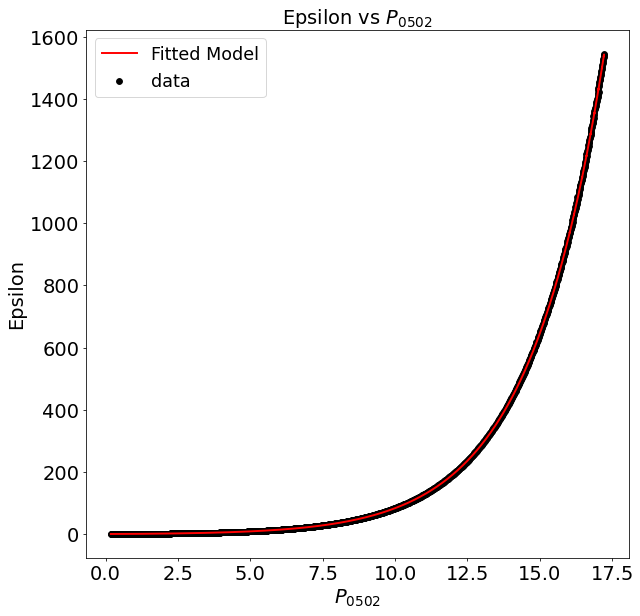}
\caption{
\label{epsilon_vs_P0502}
Approximation of $eta$ using the Petrosian concentration index $P_{0502}$ for $0.5 \leq n \leq 15$.}
\end{figure}

Though it is difficult to approximate the corrected $\epsilon$ value using the uncorrected Concentration Index ($C_{2080}$), we can use the Petrosian concentration index $P_{0502}$ (that is an uncorrected quantity, see Section \ref{petrosianconcentrationindex}) to estimate $\epsilon$. We use a combination of a $6^{th}$ degree Polynomial and a single exponential for this approximation. For clarity we denote the Petrosian concentration index $P_{0502}$ with $P$ (see Figure \ref{epsilon_vs_P0502}):

\begin{equation}\label{eq:p_vs_epsilon_approx}
 \begin{array}{l}
    \epsilon(P) \approx 1.09339566 - 0.14524911 \cdot P + 0.50361697 \cdot P^2 - 0.1215809 \cdot P^3 + 0.02533795 \cdot P^4\\
     - 0.00196243 \cdot P^5 + 0.00009081 \cdot P^6 + 0.03312881 \cdot exp\left\{ \frac{P}{1.83616642} \right\}
\end{array}
\end{equation}

This approximation is useful because the radii used to compute $P_{0502}$ do not depend on $\epsilon$ itself. This means that $P_{0502}$ can be computed directly from the data (no need for corrections) and its value can be used to compute $\epsilon$ without the need to know any of the radii that enclose a fraction of the total flux (i.e $r_{20}, r_{80}, r_{50}, r_{90}$, etc.). 

\subsubsection{Sérsic Effective and Petrosian Half-Light Radii}

\cite{grahamPetrosian2005AJ} gives the following approximation for $r_e$ using the uncorrected $r_{50}$ and $r_{90}/r_{50}$ (for Sérsic profiles with $0.1 \leq n \leq 10$ ):

\begin{equation}
    r_e \approx \frac{r_{50}}{ \left[1 - (8 \times 10^{-6}) (r_{90}/r_{50})^{8.47}\right]}
\end{equation}

\subsection{Recovering Total Sérsic Magnitudes}

Because Petrosian profiles depend on the morphology of galaxies, the Petrosian total flux radius computed using the default parameters ($r_{total} = r(\eta_{0.2}) \cdot 2$) often results in an aperture radius that underestimates the total Sérsic magnitudes (see Section \ref{petrosiancorrections}). \cite{grahamPetrosian2005AJ} discusses in detail the relationship between morphology, total Sérsic magnitudes, and Petrosian profiles. Using Equation \ref{eq:fluxenclosed},  we can define the total Sérsic magnitude ($m_{s}$) and the total Petrosian magnitude ($m_{p}$) as a function of the Petrosian radius ($r_{petro}$) and $\epsilon$ as follows: 

\begin{equation}
m_{s}(\leq r) = - 2.5 \cdot log_{10}( L(\leq r) )
\end{equation}

\begin{equation}
m_{p}(\leq \epsilon \cdot r_{petro}) = - 2.5 \cdot log_{10}( L(\leq \epsilon \cdot r_{petro}) )
\end{equation}

Using the uncorrected $r_{50}$ and $r_{90}/r_{50}$, \cite{grahamPetrosian2005AJ} (their Equation 6) provide the following approximation for the ``missing" magnitude (for Sérsic profiles with $0.1 \leq n \leq 10$):

\begin{equation}
\Delta m \equiv m_{p} - m_{s} = P_{1} \cdot exp\left[(r_{90}/r_{50})^{P_{2}}\right]
\end{equation}

\begin{figure}[h]
\centering

\includegraphics[width=16cm]{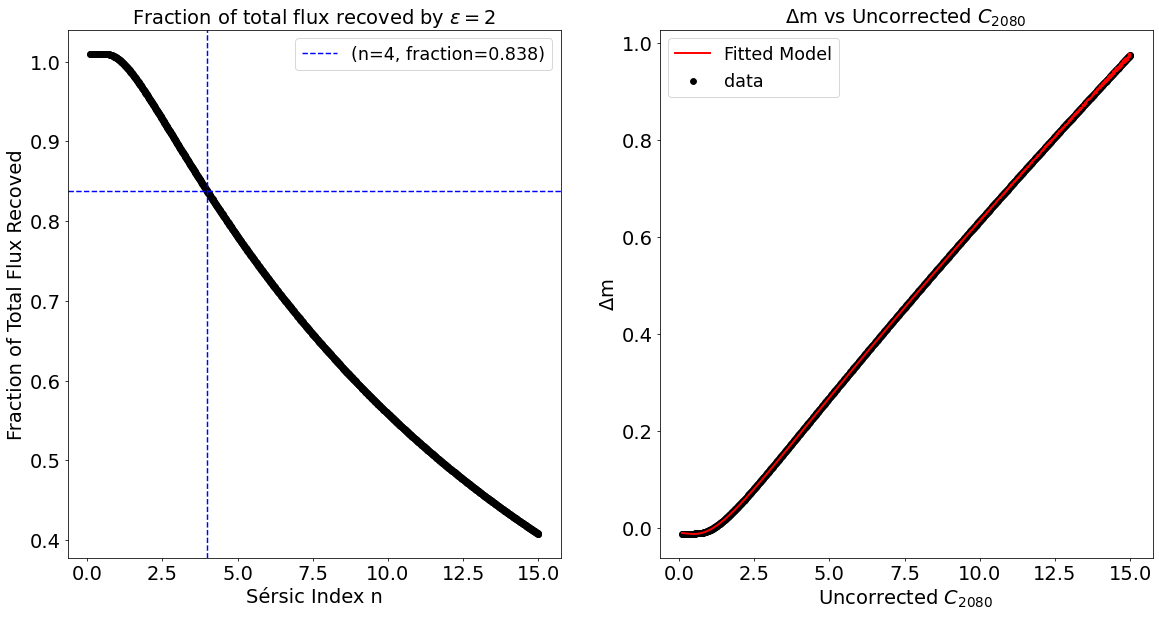}
\caption{
\label{missing_flux_plot}
The first panel shows the fraction of the total corrected flux that is recovered by using the default epsilon value ($\epsilon = 2$). The second panel shows a line of best fit that approximates the difference in corrected and uncorrected magnitudes (i.e $\Delta m$) for $0.5 \leq n \leq 15$.}
\end{figure}

Where $P_{1}$ and $P_{2}$ equal to $5.1 \cdot 10^{-4}$ and $1.451$. We build on this work and extend the domain of the approximation to $0.1 \leq n \leq 15$ using a combination of an exponential and a polynomial. We provide this approximation as a function of uncorrected $C_{2080}$ ($U$) that estimates the ``missing" magnitudes within 0.01 mags (see Figure \ref{missing_flux_plot}):

\begin{equation}
\begin{array}{l}
    \Delta m \approx -0.04990413 + 0.07810246 \cdot U - 0.04810471 \cdot U^2 \\
    + 0.00929017 \cdot U^3  + 3.241416  \cdot  10^{-6} \cdot exp\left\{ \frac{U}{0.519920} \right\}
\end{array}
\end{equation}

\subsection{Effects of PSFs on Approximations}\label{effectsofpsfapprox}

\begin{figure}[h]

\centering

\includegraphics[width=\linewidth]{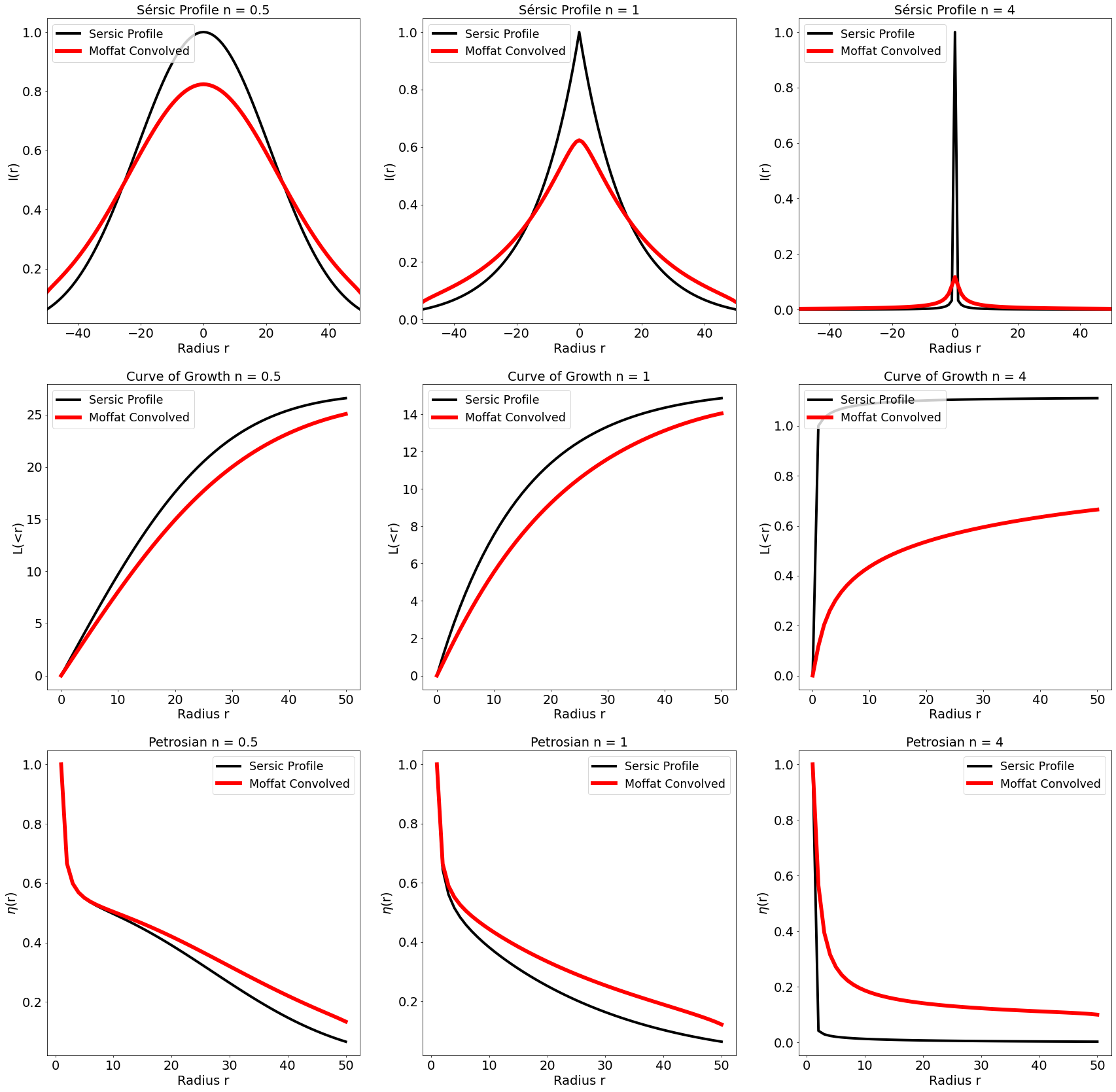}

\caption{\label{fig:psf_effect_sersic}
Three 1D Sérsic profiles are defined with $r_e = 25$ but with varying Sérsic indices. Each of the three profiles is convolved with the same PSF modeled by a Moffat function. Notice how the de Vaucouleurs' profile ($n = 4$) is most affected by the PSF convolution.}
\end{figure}

Point spread functions (PSFs) and seeing effects greatly diminish our ability to approximate Sérsic profile parameters and Sérsic radii. This is because these optical effects smear the light coming from astronomical sources causing a change in their intensity profiles (see Figure \ref{fig:psf_effect_sersic}) and curves of growth. For example, the corrected half-light radius is not equal to the Sérsic half-light radius if the profile is convolved with a PSF (which stretches out the profile). Since each PSF smears the intensity profiles of galaxies in a unique way (see Figures \ref{fig:concentrationvsepsilon} and \ref{fig:concentrationvsepsilonPSF}), a correction unique to the PSF must be applied to reconstruct the Sérsic radii. This adds an extra layer to error corrections and is the reason why we must generate a grid, using a known PSF, that maps the smeared uncorrected/corrected measurements to the intrinsic Sérsic parameters and Sérsic profile derived quantities.

\clearpage

\section{Morphology and Fitting Comparisons} \label{apdx:comparison}

Here we perform quantitative tests to compare and demonstrate the advantages of PetroFit \citep{petrofit}(v0.3.0), statmorph (v0.4.0), and GALFIT (v3.0.5). We chose statmorph for this comparison because it is an Astropy and PhotUtils based package that can produce morphological diagnoses that can be compared to PetroFit's radial and concentration measurements. For fitting, GALFIT was the best package to compare to because it has been extensively tested and is widely adopted. We run two kinds of tests, one on morphology-related measurements and a second on fitting. For the morphology tests, we ran PetroFit and statmorph on ideal Sérsic profiles (convolved with a Moffat PSFs) which were simulated using GALFIT. For the fitting test, we ran statmorph and GALFIT on the dataset discussed in Section \ref{singlesection} to see if we could reproduce similar results. To limit the tests to the fitting or morphology parameters, we provided statmorph the same segmentation maps as PetroFit and used the same background-subtracted images as input for all three packages. Further details are discussed in the following subsections. 

\subsection{Fitting Test}

As a demonstration of PetroFit's fitting libraries, we compare the results from Section \ref{singlefitting} to fits we perform using GALFIT and statmorph. We provided both GALFIT and statmorph with the same background-subtracted image used by PetroFit. For statmorph, we follow the same segmentation step as used by PetroFit. We provided GALFIT with the same source mask used by PetroFit. We also provided GALFIT with parameters estimates close to those of PetroFit and statmorph (after fitting using the later packages). To make fair timing comparisons, we consider run-times that include loading the data from file and refer to it as ``total run-time". We do not include segmentation time since all three packages use maps and masks from the same segmentation steps. Running on an Intel Core i7-2600 CPU at 3.40GHz,  PetroFit finished with a total run-time under 1.2 seconds, statmorph under 0.8 seconds, and GALFIT under 1.8 seconds. The fit results are given in Table \ref{table:fttingcomp}. Radii are given in pixels while photometric flux measurements are given in electrons per second (except magnitudes, which are given in the AB system). The results show similar outputs as expected. PetroFit and statmorph produced similar models and upon visual inspection, show the least amount of residual.

\begin{table}[h]
\centering
\caption{Comparison Table of Fitted Parameters}\label{table:fttingcomp}
\begin{tabular}{|p{2.5cm}|m{2cm}|m{2cm}|m{2cm}|}
\hline
Parameter & PetroFit & statmorph & GALFIT \\
\hline
amplitude & 0.087 & 0.086 & N/A\\
$r_{eff}$ & 7.737 & 7.777 & 7.81\\
n  &  1.62 & 1.62 & 1.5212 \\
$x_0$ &  75.906 & 74.906 & 75.4388\\
$y_0$ & 75.408 & 75.41 & 75.1472\\
ellip & 0.109 &  0.107 & 0.10\\
theta & -1.136 & 2.009 & 0.65\\
psf\_pa & 0.00 & N/A & N/A\\  
Total Mag & 21.76 & 21.77 & 21.68 \\ 
Total Time (s) & 1.14 & 0.76 & 1.71 \\
\hline
\end{tabular}
\end{table}

\subsection{Morphology Test}

For this test, we first generated mock galaxies using single component Sérsic profiles and an ideal Moffat PSF. The Sérsic profiles shared the same parameters ($r_{eff}=15$ pix, $ellip=0$, $\theta=0$), except for the Sérsic index and effective intensity. The Sérsic index was set to (0.5, 1, 4, or 5) and the effective intensity was defined such that the total Sérsic flux of the profile was equal to one (total flux equal to 0.99 by PetroFit definition, see Section \ref{totalfluxdef}). The Moffat profile was generated by first fitting a real Hubble F105W PSF shown in Figure \ref{fig:singlegalaxy} and converting that model into an image. This was done so we could create a PSF with no noise and exactly centered in the image. Using the model parameters and the Moffat PSF, we used GALFIT to create model images of size $900$ on each axis. A correction grid was generated using the new PSF for PetroFit. We carried out the segmentation and deblending step using the $make_catalog$ function. This gave us PhotUtils segmentation maps that we used for both tools (both tools used the same maps). We ran statmorph using its \texttt{SourceMorphology} class. We placed PetroFit's photometry, morphology, and fitting steps in a single function for this test. For profiles with a Sérsic index greater than three, PetroFit oversampled the model by a factor of $10$ within a box of $50$ by $50$ pixels of the center. Both functions computed morphology parameters and fit the input images using the Moffat PSF. 

The following tables show the results of the tests. We show the results from each profile in a separate table. The ``Values" column shows the parameter name and the ``GALFIT Inputs" column shows Sérsic parameters that were used to create the image. Values such as $r_{20}$ and ``Total Flux" in the ``GALFIT Inputs" column were calculated using analytical equations defined in Sections \ref{sec:intro} and \ref{sec:photo}. It is important to note that these ideal values are proxies since they do not account for PSF smearing (see Appendix \ref{effectsofpsfapprox}), thus radii close to the center (such as $r_{20}$) will be skewed. Lastly, each table under the ``PetroFit" and ``statmorph" columns show the run-time of each tool in seconds (running on an Intel Core i7-2600 CPU at 3.40GHz). Radii are given in pixels while photometric flux measurements are given in electrons per second. The run-times show similar run-times for this sample, with PetroFit performing better with high index sources. As stated in Section \ref{sec:otherpackages}, statmorph computes additional parameters, such as Gini-M20 statistics, which contribute to its run time. PetroFit's correction grids allow it to get much better estimates of $r_{80}$ and $C_{2080}$ for highly concentrated profiles (see $n= 4$ and $5$). For example, for a profile with $n=4$ statmorph measures a total flux of $0.83519\ e^-/s$ which is $~84\%$ of the total Sérsic flux (this is expected, see Figure \ref{missing_flux_plot}). It's also important to note that PetroFit was able to predict these values before fitting the images. Another important result is that oversampling the high index images allowed PetroFit to produce better fits for $n$. 

\begin{table}[h]
\centering
\caption{Gaussian Profile $n=0.5$}
\begin{filecontents*}{galfit_simulated_sersic_n0.5_psf_comparison.csv}
Values,GALFIT Inputs,PetroFit,statmorph
amplitude,0.0004903,0.00049024,0.00049019
$r_{eff}$,15.0,15.001,15.002
n,0.5,0.49977,0.4999
$x_0$,450.0,450.0,450.0
$y_0$,450.0,450.0,450.0
ellip,0.0,9.1459e-07,1.877e-05
theta,0.0,1.5676,1.4208
$r_{petro}$,29.39,30.606,29.896
$r_{20}$,8.5108,8.6417,8.6407
$r_{80}$,22.857,22.955,23.181
$C_{2080}$,2.1452,2.1213,2.1429
Total Flux,0.99,0.98875,0.99987
Time,0.0,9.8942,4.1861
\end{filecontents*}
\csvautotabular{galfit_simulated_sersic_n0.5_psf_comparison.csv}
\end{table}

\begin{table}[h]
\centering
\caption{Exponential Profile $n=1$}
\begin{filecontents*}{galfit_simulated_sersic_n1_psf_comparison.csv}
Values,GALFIT Inputs,PetroFit,statmorph
amplitude,0.00037197,0.00037223,0.00037173
$r_{eff}$,15.0,14.998,15.005
n,1.0,0.99749,0.99936
$x_0$,450.0,450.0,450.0
$y_0$,450.0,450.0,450.0
ellip,0.0,0.0,9.4263e-07
theta,0.0,-6.2832,1.507
$r_{petro}$,32.37,33.847,32.888
$r_{20}$,7.3679,7.5615,7.5143
$r_{80}$,26.761,26.555,25.874
$C_{2080}$,2.8008,2.7277,2.6849
Total Flux,0.99,0.98838,0.99431
Time,0.0,10.912,8.7325
\end{filecontents*}
\csvautotabular{galfit_simulated_sersic_n1_psf_comparison.csv}
\end{table}

\begin{table}[h]
\centering
\caption{De Vaucouleurs' Profile $n=4$}
\begin{filecontents*}{galfit_simulated_sersic_n4_psf_comparison.csv}
Values,GALFIT Inputs,PetroFit,statmorph
amplitude,0.00019609,0.0001959,0.0001858
$r_{eff}$,15.0,15.004,15.442
n,4.0,4.0046,3.8692
$x_0$,450.0,450.0,450.11
$y_0$,450.0,450.0,450.11
ellip,0.0,0.0,0.00031198
theta,0.0,-6.2832,2.3305
$r_{petro}$,27.31,29.886,28.357
$r_{20}$,4.1917,4.951,3.9814
$r_{80}$,47.535,45.459,23.376
$C_{2080}$,5.2731,4.8146,3.8437
Total Flux,0.99,0.98651,0.83519
Time,0.0,12.246,31.045
\end{filecontents*}
\csvautotabular{galfit_simulated_sersic_n4_psf_comparison.csv}
\end{table}

\begin{table}[h]
\centering
\caption{High Index Profile $n=5$}
\begin{filecontents*}{galfit_simulated_sersic_n5_psf_comparison.csv}
Values,GALFIT Inputs,PetroFit,statmorph
amplitude,0.000176,0.00017547,0.00016352
$r_{eff}$,15.0,15.019,15.598
n,5.0,5.0149,4.787
$x_0$,450.0,450.0,449.88
$y_0$,450.0,450.0,449.88
ellip,0.0,0.0,-0.00017608
theta,0.0,-6.2832,0.37224
$r_{petro}$,23.78,26.735,25.157
$r_{20}$,3.653,4.5009,3.4316
$r_{80}$,54.582,51.58,20.844
$C_{2080}$,5.872,5.2959,3.9174
Total Flux,0.99,0.985,0.78326
Time,0.0,13.446,32.105
\end{filecontents*}
\csvautotabular{galfit_simulated_sersic_n5_psf_comparison.csv}
\end{table}

%% \section{Sampling Galaxies onto Pixels} %% not enough time 

\newpage
\bibliography{petrofit}{}

\begin{thebibliography}{}
\expandafter\ifx\csname natexlab\endcsname\relax\def\natexlab#1{#1}\fi
\providecommand{\url}[1]{\href{#1}{#1}}
\providecommand{\dodoi}[1]{doi:~\href{http://doi.org/#1}{\nolinkurl{#1}}}
\providecommand{\doeprint}[1]{\href{http://ascl.net/#1}{\nolinkurl{http://ascl.net/#1}}}
\providecommand{\doarXiv}[1]{\href{https://arxiv.org/abs/#1}{\nolinkurl{https://arxiv.org/abs/#1}}}

\bibitem[{{Andredakis} {et~al.}(1995){Andredakis}, {Peletier}, \&
  {Balcells}}]{andredakis1995MNRAS}
{Andredakis}, Y.~C., {Peletier}, R.~F., \& {Balcells}, M. 1995, \mnras, 275,
  874, \dodoi{10.1093/mnras/275.3.874}

\bibitem[{{Astropy Collaboration} {et~al.}(2013){Astropy Collaboration},
  {Robitaille}, {Tollerud}, {Greenfield}, {Droettboom}, {Bray}, {Aldcroft},
  {Davis}, {Ginsburg}, {Price-Whelan}, {Kerzendorf}, {Conley}, {Crighton},
  {Barbary}, {Muna}, {Ferguson}, {Grollier}, {Parikh}, {Nair}, {Unther},
  {Deil}, {Woillez}, {Conseil}, {Kramer}, {Turner}, {Singer}, {Fox}, {Weaver},
  {Zabalza}, {Edwards}, {Azalee Bostroem}, {Burke}, {Casey}, {Crawford},
  {Dencheva}, {Ely}, {Jenness}, {Labrie}, {Lim}, {Pierfederici}, {Pontzen},
  {Ptak}, {Refsdal}, {Servillat}, \& {Streicher}}]{astropy13}
{Astropy Collaboration}, {Robitaille}, T.~P., {Tollerud}, E.~J., {et~al.} 2013,
  \aap, 558, A33, \dodoi{10.1051/0004-6361/201322068}

\bibitem[{{Bershady}(1995)}]{Bershady1995AJ}
{Bershady}, M.~A. 1995, \aj, 109, 87, \dodoi{10.1086/117259}

\bibitem[{{Bershady} {et~al.}(2000){Bershady}, {Jangren}, \&
  {Conselice}}]{bershady2000AJ}
{Bershady}, M.~A., {Jangren}, A., \& {Conselice}, C.~J. 2000, \aj, 119, 2645,
  \dodoi{10.1086/301386}

\bibitem[{{Bertin} \& {Arnouts}(1996)}]{sextractor1996}
{Bertin}, E., \& {Arnouts}, S. 1996, \aaps, 117, 393,
  \dodoi{10.1051/aas:1996164}

\bibitem[{{Blanton} {et~al.}(2001){Blanton}, {Dalcanton}, {Eisenstein},
  {Loveday}, {Strauss}, {SubbaRao}, {Weinberg}, {Anderson}, {Annis}, {Bahcall},
  {Bernardi}, {Brinkmann}, {Brunner}, {Burles}, {Carey}, {Castander},
  {Connolly}, {Csabai}, {Doi}, {Finkbeiner}, {Friedman}, {Frieman}, {Fukugita},
  {Gunn}, {Hennessy}, {Hindsley}, {Hogg}, {Ichikawa}, {Ivezi{\'c}}, {Kent},
  {Knapp}, {Lamb}, {Leger}, {Long}, {Lupton}, {McKay}, {Meiksin}, {Merelli},
  {Munn}, {Narayanan}, {Newcomb}, {Nichol}, {Okamura}, {Owen}, {Pier}, {Pope},
  {Postman}, {Quinn}, {Rockosi}, {Schlegel}, {Schneider}, {Shimasaku},
  {Siegmund}, {Smee}, {Snir}, {Stoughton}, {Stubbs}, {Szalay}, {Szokoly},
  {Thakar}, {Tremonti}, {Tucker}, {Uomoto}, {Vanden Berk}, {Vogeley},
  {Waddell}, {Yanny}, {Yasuda}, \& {York}}]{blanton2001AJ}
{Blanton}, M.~R., {Dalcanton}, J., {Eisenstein}, D., {et~al.} 2001, \aj, 121,
  2358, \dodoi{10.1086/320405}

\bibitem[{Bradley {et~al.}(2020)Bradley, Sipőcz, Robitaille, Tollerud,
  Vinícius, Deil, Barbary, Wilson, Busko, Günther, Cara, Conseil, Bostroem,
  Droettboom, Bray, Bratholm, Lim, Barentsen, Craig, Pascual, Perren, Greco,
  Donath, de~Val-Borro, Kerzendorf, Bach, Weaver, D'Eugenio, Souchereau, \&
  Ferreira}]{bradley2020}
Bradley, L., Sipőcz, B., Robitaille, T., {et~al.} 2020, astropy/photutils:
  1.0.0, 1.0.0,  Zenodo, \dodoi{10.5281/zenodo.4044744}

\bibitem[{{Ciotti}(1991)}]{Ciotti1991}
{Ciotti}, L. 1991, \aap, 249, 99

\bibitem[{{Conselice}(2003)}]{Conselice2003ApJS}
{Conselice}, C.~J. 2003, \apjs, 147, 1, \dodoi{10.1086/375001}

\bibitem[{{Crawford}(2006)}]{crawford2006PhDT}
{Crawford}, S.~M. 2006, PhD thesis, The University of Wisconsin - Madison,
  Wisconsin, USA

\bibitem[{{de Vaucouleurs}(1948)}]{vaucouleurs1948}
{de Vaucouleurs}, G. 1948, Annales d'Astrophysique, 11, 247

\bibitem[{{Ferrari} {et~al.}(2004){Ferrari}, {Dottori}, {Caon}, {Nobrega}, \&
  {Pavani}}]{ferrari2004}
{Ferrari}, F., {Dottori}, H., {Caon}, N., {Nobrega}, A., \& {Pavani}, D.~B.
  2004, \mnras, 347, 824, \dodoi{10.1111/j.1365-2966.2004.07254.x}

\bibitem[{Geda {et~al.}(2022{\natexlab{a}})Geda, Crawford, Hunt, Bershady,
  Tollerud, \& Randriamampandry}]{petrofitpapers}
Geda, R., Crawford, S.~M., Hunt, L., {et~al.} 2022{\natexlab{a}},
  PetroFit/petrofit\_papers: Version 1.0, v1.0,  Zenodo,
  \dodoi{10.5281/zenodo.6040890}

\bibitem[{Geda {et~al.}(2022{\natexlab{b}})Geda, Crawford, Hunt, Bershady,
  Tollerud, \& Randriamampandry}]{petrofit}
---. 2022{\natexlab{b}}, PetroFit/petrofit: Version 0.3, v0.3,  Zenodo,
  \dodoi{10.5281/zenodo.6040781}

\bibitem[{{Graham} {et~al.}(1996){Graham}, {Lauer}, {Colless}, \&
  {Postman}}]{graham1996ApJ}
{Graham}, A., {Lauer}, T.~R., {Colless}, M., \& {Postman}, M. 1996, \apj, 465,
  534, \dodoi{10.1086/177440}

\bibitem[{{Graham}(2001)}]{graham2001AJ}
{Graham}, A.~W. 2001, \aj, 121, 820, \dodoi{10.1086/318767}

\bibitem[{{Graham} \& {Driver}(2005)}]{graham2005PASA}
{Graham}, A.~W., \& {Driver}, S.~P. 2005, \pasa, 22, 118,
  \dodoi{10.1071/AS05001}

\bibitem[{{Graham} {et~al.}(2005){Graham}, {Driver}, {Petrosian}, {Conselice},
  {Bershady}, {Crawford}, \& {Goto}}]{grahamPetrosian2005AJ}
{Graham}, A.~W., {Driver}, S.~P., {Petrosian}, V., {et~al.} 2005, \aj, 130,
  1535, \dodoi{10.1086/444475}

\bibitem[{{Graham} {et~al.}(2001){Graham}, {Erwin}, {Caon}, \&
  {Trujillo}}]{grahamgrwincaon2001ApJ}
{Graham}, A.~W., {Erwin}, P., {Caon}, N., \& {Trujillo}, I. 2001, \apjl, 563,
  L11, \dodoi{10.1086/338500}

\bibitem[{Harris {et~al.}(2020)Harris, Millman, van~der Walt, Gommers,
  Virtanen, Cournapeau, Wieser, Taylor, Berg, Smith, Kern, Picus, Hoyer, van
  Kerkwijk, Brett, Haldane, Fernández~del Río, Wiebe, Peterson,
  Gérard-Marchant, Sheppard, Reddy, Weckesser, Abbasi, Gohlke, \&
  Oliphant}]{2020NumPy-Array}
Harris, C.~R., Millman, K.~J., van~der Walt, S.~J., {et~al.} 2020, Nature, 585,
  357–362, \dodoi{10.1038/s41586-020-2649-2}

\bibitem[{Hunter(2007)}]{Matplotlib}
Hunter, J.~D. 2007, Computing in Science Engineering, 9, 90,
  \dodoi{10.1109/MCSE.2007.55}

\bibitem[{{Kent}(1985)}]{kent1985ApJS}
{Kent}, S.~M. 1985, \apjs, 59, 115, \dodoi{10.1086/191066}

\bibitem[{{Khosroshahi} {et~al.}(2000){Khosroshahi}, {Wadadekar}, \&
  {Kembhavi}}]{khosroshahi2000ApJ}
{Khosroshahi}, H.~G., {Wadadekar}, Y., \& {Kembhavi}, A. 2000, \apj, 533, 162,
  \dodoi{10.1086/308654}

\bibitem[{{Kron}(1980)}]{kron80}
{Kron}, R.~G. 1980, \apjs, 43, 305, \dodoi{10.1086/190669}

\bibitem[{Levenberg(1944)}]{levenberg1944AMF}
Levenberg, K. 1944, Quarterly of Applied Mathematics, 2, 164

\bibitem[{{Lotz} {et~al.}(2017){Lotz}, {Koekemoer}, {Coe}, {Grogin}, {Capak},
  {Mack}, {Anderson}, {Avila}, {Barker}, {Borncamp}, {Brammer}, {Durbin},
  {Gunning}, {Hilbert}, {Jenkner}, {Khandrika}, {Levay}, {Lucas}, {MacKenty},
  {Ogaz}, {Porterfield}, {Reid}, {Robberto}, {Royle}, {Smith},
  {Storrie-Lombardi}, {Sunnquist}, {Surace}, {Taylor}, {Williams}, {Bullock},
  {Dickinson}, {Finkelstein}, {Natarajan}, {Richard}, {Robertson}, {Tumlinson},
  {Zitrin}, {Flanagan}, {Sembach}, {Soifer}, \& {Mountain}}]{Lotz2017ApJ}
{Lotz}, J.~M., {Koekemoer}, A., {Coe}, D., {et~al.} 2017, \apj, 837, 97,
  \dodoi{10.3847/1538-4357/837/1/97}

\bibitem[{Marquardt(1963)}]{marquardt1963AnAF}
Marquardt, D. 1963, Journal of The Society for Industrial and Applied
  Mathematics, 11, 431

\bibitem[{{Merlin} {et~al.}(2016){Merlin}, {Amor{\'\i}n}, {Castellano},
  {Fontana}, {Buitrago}, {Dunlop}, {Elbaz}, {Boucaud}, {Bourne}, {Boutsia},
  {Brammer}, {Bruce}, {Capak}, {Cappelluti}, {Ciesla}, {Comastri}, {Cullen},
  {Derriere}, {Faber}, {Ferguson}, {Giallongo}, {Grazian}, {Lotz},
  {Micha{\l}owski}, {Paris}, {Pentericci}, {Pilo}, {Santini}, {Schreiber},
  {Shu}, \& {Wang}}]{astrodeep2016A&A}
{Merlin}, E., {Amor{\'\i}n}, R., {Castellano}, M., {et~al.} 2016, \aap, 590,
  A30, \dodoi{10.1051/0004-6361/201527513}

\bibitem[{{M{\"o}llenhoff} \& {Heidt}(2001)}]{mollenhoff2001AA}
{M{\"o}llenhoff}, C., \& {Heidt}, J. 2001, \aap, 368, 16,
  \dodoi{10.1051/0004-6361:20000335}

\bibitem[{Mor{\'{e}} {et~al.}(1980)Mor{\'{e}}, Garbow, \& Hillstrom}]{minpack}
Mor{\'{e}}, J.~J., Garbow, B.~S., \& Hillstrom, K.~E. 1980, User Guide for
  {MINPACK-1}, Tech. Rep. ANL-80-74, Argonne National Laboratory, Argonne, IL,
  USA

\bibitem[{{Moriondo} {et~al.}(1998){Moriondo}, {Giovanardi}, \&
  {Hunt}}]{moriondo1998AA}
{Moriondo}, G., {Giovanardi}, C., \& {Hunt}, L.~K. 1998, \aap, 339, 409.
\newblock \doarXiv{astro-ph/9802140}

\bibitem[{{Peng} {et~al.}(2010){Peng}, {Ho}, {Impey}, \& {Rix}}]{peng2010}
{Peng}, C.~Y., {Ho}, L.~C., {Impey}, C.~D., \& {Rix}, H.-W. 2010, \aj, 139,
  2097, \dodoi{10.1088/0004-6256/139/6/2097}

\bibitem[{{Petrosian}(1976)}]{petrosian1976ApJ}
{Petrosian}, V. 1976, \apjl, 210, L53, \dodoi{10.1086/182301}

\bibitem[{{Rodriguez-Gomez} {et~al.}(2019){Rodriguez-Gomez}, {Snyder}, {Lotz},
  {Nelson}, {Pillepich}, {Springel}, {Genel}, {Weinberger}, {Tacchella},
  {Pakmor}, {Torrey}, {Marinacci}, {Vogelsberger}, {Hernquist}, \&
  {Thilker}}]{rodriguezGomez2019MNRAS}
{Rodriguez-Gomez}, V., {Snyder}, G.~F., {Lotz}, J.~M., {et~al.} 2019, \mnras,
  483, 4140, \dodoi{10.1093/mnras/sty3345}

\bibitem[{{S{\'e}rsic}(1963)}]{sersic1963}
{S{\'e}rsic}, J.~L. 1963, Boletin de la Asociacion Argentina de Astronomia La
  Plata Argentina, 6, 41

\bibitem[{{S{\'e}rsic}(1968)}]{sersic1968}
---. 1968, {Atlas de Galaxias Australes}

\bibitem[{{Strauss} {et~al.}(2002){Strauss}, {Weinberg}, {Lupton}, {Narayanan},
  {Annis}, {Bernardi}, {Blanton}, {Burles}, {Connolly}, {Dalcanton}, {Doi},
  {Eisenstein}, {Frieman}, {Fukugita}, {Gunn}, {Ivezi{\'c}}, {Kent}, {Kim},
  {Knapp}, {Kron}, {Munn}, {Newberg}, {Nichol}, {Okamura}, {Quinn}, {Richmond},
  {Schlegel}, {Shimasaku}, {SubbaRao}, {Szalay}, {Vanden Berk}, {Vogeley},
  {Yanny}, {Yasuda}, {York}, \& {Zehavi}}]{strauss2002AJ}
{Strauss}, M.~A., {Weinberg}, D.~H., {Lupton}, R.~H., {et~al.} 2002, \aj, 124,
  1810, \dodoi{10.1086/342343}

\bibitem[{{Trujillo} {et~al.}(2001){Trujillo}, {Graham}, \&
  {Caon}}]{trujillo2001}
{Trujillo}, I., {Graham}, A.~W., \& {Caon}, N. 2001, \mnras, 326, 869,
  \dodoi{10.1046/j.1365-8711.2001.04471.x}

\bibitem[{Virtanen {et~al.}(2020)Virtanen, Gommers, Oliphant, Haberland, Reddy,
  Cournapeau, Burovski, Peterson, Weckesser, Bright, {van der Walt}, Brett,
  Wilson, Millman, Mayorov, Nelson, Jones, Kern, Larson, Carey, Polat, Feng,
  Moore, {VanderPlas}, Laxalde, Perktold, Cimrman, Henriksen, Quintero, Harris,
  Archibald, Ribeiro, Pedregosa, {van Mulbregt}, \& {SciPy 1.0
  Contributors}}]{2020SciPy-NMeth}
Virtanen, P., Gommers, R., Oliphant, T.~E., {et~al.} 2020, Nature Methods, 17,
  261, \dodoi{10.1038/s41592-019-0686-2}

\end{thebibliography}
\bibliographystyle{aasjournal}



\end{document}